\documentclass[amssymb,twocolumn,aps]{revtex4}
\usepackage{graphics,color,graphicx,amsmath}
\usepackage[monochrome]{xcolor}
\usepackage{subcaption}
\usepackage{natbib}
\usepackage{verbatim}
\usepackage{graphicx}
\usepackage[above,below]{placeins}
\usepackage{float}
\usepackage{tabularx}
\usepackage{array}
\usepackage{times}
\usepackage{blindtext}
\usepackage{url}
\usepackage{rotating}
\usepackage{mathtools}
\usepackage{threeparttable}
\usepackage{enumitem}

\setlist{nosep}                 

\begin{document}
  
  \title{Characterizing adopters of named active learning methods in physics: An interview study}
 
  \author{Ibukunoluwa Bukola~\footnote[1]{ib396@drexel.edu}$^{1}$, Eric Brewe~\footnote[2]{Contact author: eb573@drexel.edu}$^1$, Adrienne L. Traxler$^{2}$, and Meagan Sundstrom$^{1}$}
    \affiliation{$^{1}$Department of Physics, Drexel University, Philadelphia, Pennsylvania 19104, USA\\
    $^{2}$Department of Science Education, University of Copenhagen, Copenhagen, Denmark}
 

  \begin{abstract}
Active learning has been shown to be more beneficial for student learning than didactic lecturing. Consequently, researchers have developed a slate of active learning strategies for use in physics instruction, including course-level reforms that are widely known by specific names. To increase adoption and support prolonged implementation of these named active learning methods, it is important to understand the backgrounds and experiences of instructors who are already using them. In this study, we interviewed 20 introductory physics and astronomy instructors who use one of Student-Centered Active Learning Environment with Upside-Down Pedagogies (SCALE-UP), Investigative Science Learning Environment (ISLE), Tutorials in Introductory Physics and Astronomy, or Peer Instruction to gain insight into their teaching experiences, professional development, pathways to learning about their method, and supports for and difficulties in their implementation. We find for our sample that all implementers are involved with the Discipline-Based Education Research (DBER) community (e.g., conducting DBER work, engaging with DBER researchers or content, receiving training for evidence-based instructional practices), but there is variation in the extent of engagement with DBER across instructors. While instructors identify a variety of supports for their implementation, such as from the broader education community and departmental or institutional administration, the difficulties they experience are representative of both long-standing issues associated with active learning adoption (e.g., physical classroom and timing constraints, student resistance) and new issues related to emergent technologies (e.g., Artificial Intelligence). Finally, we uncover differences in implementers across the four named methods, particularly related to their training for the method and the supports they received for implementation. Together, these findings underscore the importance of supporting multiple dissemination avenues for named active learning methods, point to a possible need for more training opportunities for SCALE-UP and ISLE in particular, and identify nascent issues in the contemporary active learning classroom that future work may further characterize and address.  
  \end{abstract}

  \maketitle

\section{Introduction}
\label{intro}

The use of active learning in undergraduate science instruction has well-documented benefits to students across a range of outcomes, including conceptual learning, self-efficacy, and retention~\cite{freeman2014active, theobald2020active, bullard2008effects, salomone2017required, stump2014engineering, rossi2021active}. Over the past decade, instructors have increasingly implemented active learning, including adoption of these strategies across different educational levels and geographic locations~\cite{sokoloff2021active, hake1998interactive, dancy2016faculty, dancy2024physics, stains2018anatomy}. Despite the growing use of active learning, traditional lecturing is still common~\cite{dancy2024physics, stains2018anatomy}. As a result, there have been widespread calls for instructors to pivot from more traditional, didactic ways of teaching to active learning strategies~\cite{lumpkin2015student, freeman2014active, olson2012engage}.  

To promote adoption of active learning, research has started to unpack instructor pathways to using active learning strategies and the supports for and barriers to implementation (e.g., Refs.~\cite{dancy2016faculty, sokoloff2021active, foote2014diffusion,west2022stem, borte2023barriers, brownell2012barriers}). This body of work, however, is sometimes limited to single institutions, single courses, and/or single active learning strategies, making it difficult to draw broad conclusions about how and why active learning adoption occurs~\cite{dancy2016faculty,foote2014diffusion, affriyenni2025navigating, hall2002adoption, turpen2007understanding, givens2020faculty}. In other cases, studies span many instructors and implementations, but lack sufficient detail about the institutional contexts and active learning strategies being used to devise targeted implications for dissemination and adoption~\cite{chasteen2015educational, west2022stem, affriyenni2025navigating, funk2022role}. Studies that both capture a range of institutional contexts and implementations and contain sufficient information about the instructors and their contexts are needed to develop a more systematic understanding of the diffusion of active learning. 

In physics, named active learning methods offer useful case studies for this type of work. Physics education researchers have developed a range of active learning techniques over the past three decades, from single clicker questions to unit-length interventions to these course-level reforms that are known by specific names (e.g., Peer Instruction~\cite{mazur1997peer}). Such named methods are rigorously validated by the developers, often through quasi-experimental procedures, and have demonstrated gains in student learning~\cite{missett2015searching, henderson2012use, rahman2020evaluating}. Named methods also carry a well-defined developmental lineage (e.g., because they are often motivated by specific theories of learning), structure, and set of supporting resources (e.g., the Peer Instruction textbook~\cite{mazur1997peer}). While individual implementations of the same named method may vary~\cite{turpen2009not,sundstrom2026beyond}, the shared name, structure, and goals allow them to be compared across implementations and institutions.

In this study, therefore, we closely examine the dissemination and use of four named active learning methods in physics and astronomy instruction by conducting interviews with 20 instructors using one of these methods. Several other named active learning methods for introductory physics exist (e.g., Modeling Instruction~\cite{brewe2008modeling} and Context-Rich Problems~\cite{heller1992teaching,heller1992teaching2}), but we limit the current study to the following methods because they are both well-known and widely implemented~\cite{dancy2024physics}, and we aimed to collect data from multiple implementations of each method (the methods are ordered from most to least structural reform required): 
\begin{itemize}

    \item \textit{Student-Centered Active Learning Environment with Upside-Down Pedagogies (SCALE-UP)}~\cite{beichner1999student, beichner2000introduction, beichner2007student, beichner2008scale}: Designed as an expanded version of Studio Physics~\cite{wilson1994cuple}, SCALE-UP was developed with the goal of actively engaging students in large-enrollment physics courses. The classroom setup and management are the primary focus of SCALE-UP, rather than a specific curriculum. SCALE-UP classrooms typically contain round or triangular tables that can seat three groups of three students, where students are expected to share ideas within their groups and, sometimes, with other nearby groups. Instead of the usual physics course structure of separate lecture, recitation, and laboratory (lab) sections, students in a SCALE-UP course typically attend three two-hour long classroom meetings where they work on short hands-on activities, conceptual questions, and  lab activities. To complement activities that occur during class time, students are asked to complete readings prior to attending class meetings. 
    
    \item \textit{Investigative Science Learning Environment (ISLE)}~\cite{etkina2006using, etkina2007investigative, etkina2021investigative}: ISLE focuses on engaging students in the scientific process as they learn physics. The ISLE cycle usually requires students working in groups to observe a phenomenon, come up with multiple explanations for the patterns they observe, and then make predictions derived from their explanations for a testing experiment. Based on the results of the testing experiment, they then revise, reject, or accept their proposed explanations. The purpose of engaging students in this process is to help them construct knowledge in a manner that mirrors that of professional scientists. Scientific abilities, moreso than conceptual knowledge, are emphasized. ISLE may be implemented both in the lecture and lab components of a course or in labs only. In contrast with other named methods, students in an ISLE course are encouraged to read the assigned texts \textit{after} classroom meetings.  
    
    \item \textit{Tutorials in Introductory Physics and Astronomy}~\cite{finkelstein2005replicating, adams2003lecture, mcdermott2002tutorials}: Tutorials are sets of worksheets designed to elicit and address common student misconceptions in physics. The method emphasizes the building of both conceptual understanding and scientific reasoning before students engage in mathematical calculations. The worksheets are usually administered to students during recitations, where they work in groups to complete them. Instructors or Teaching Assistants (TAs) serve as guides for students during class time. 

    \item \textit{Peer Instruction}~\cite{mazur1997peer, mazur1997peerconf, crouch2001peer}: Peer Instruction was one of the first active learning methods to be popularized for use in physics classrooms, and the method was originally designed to promote interactive engagement in large lecture classes. Peer Instruction classes usually consist of a repeated cycle of the instructor briefly lecturing, students answering a conceptual question (often multiple choice) individually---and then in groups---using a classroom response system (e.g., clickers), and then the instructor displaying and discussing the answer. Students are usually assigned pre-class readings and take a short quiz before attending lecture to confirm that the reading was done. 
\end{itemize}



\subsection{Diffusion and implementation of named active learning methods}

Prior studies have identified several factors that influence instructors' adoption of active learning strategies broadly and named methods specifically~\cite{bathgate2019perceived, albuquerque2023awareness, lund2015importance}. The first step to adoption is initial exposure to or awareness of the method, though relatively few studies have examined how instructors first learn about the specific active learning strategy they use. One study of 35 Peer Instruction users found that discussions with colleagues is the most common means of spreading knowledge of this method, and that faculty are likely to follow up with training through more formal avenues~\cite{dancy2016faculty}. The authors also observed that instructor experience with Peer Instruction during their graduate education (e.g., while working with a Physics Education Research (PER) group) helps them implement the method later on as faculty. Another study similarly found that SCALE-UP users tend to find out about the method through interaction with colleagues or other faculty~\cite{foote2014diffusion}. 

Following exposure, factors that impact instructors' decision to use an active learning method include their instructional context and their own background and training~\cite{yik2022association, yik2022evaluating, henderson2007barriers}. Yik \textit{et al}.~\cite{yik2022association} found that teaching at institutions that value the use of these strategies and/or having classroom spaces designed to support the use of active learning facilitate faculty decisions to use these methods. Related to instructors' backgrounds and the training instructors receive for implementing active learning, some studies have found that instructors who conduct science education research are more likely to be familiar with and use active learning~\cite{andrews2011active, henderson2011facilitating}. Other studies have found that faculty who attend professional development workshops are more likely to know about and use active learning strategies than faculty who do not~\cite{walczyk2003use, brawner2002survey, henderson2008promoting}. Relatedly, other work has identified a likely relationship between the type of training instructors receive and their sustained use of the methods~\cite{wieman2013use}. This study found that for faculty who were trained by science education experts, only one out of 70 stopped using evidence-based teaching practices. Meanwhile, other work documented that roughly one-third of 722 general physics faculty stopped using research-based teaching strategies after trying them~\cite{henderson2012use}. Together, these studies indicate that institutional context, instructor training for active learning methods, and instructor experience with science education research likely impact both their initial and sustained implementation of active learning. 

There is also substantial literature on the available supports for and barriers to implementation of active learning strategies~\cite{bathgate2019perceived, bathgate2019supports, henderson2007barriers, funk2022role, borte2023barriers, wieman2013use, brownell2012barriers}. Bathgate \textit{et al.}~\cite{bathgate2019supports} found that of two types of supports instructors might have or receive---personal factors (e.g., instructor's own values and confidence) and institutional or departmental factors (e.g., supportive colleagues or access to resources)---institutional or departmental factors are more predictive of instructor use of evidence-based practices. Another study on the same set of instructors found that supports are often related to one another and that receiving even one of form support can result in a positive ripple effect by yielding further support~\cite{bathgate2019perceived}. For example, having a colleague that uses active learning strategies can lead to access to helpful resources as another form of support. Conversely, relating to barriers, Henderson and Dancy interviewed five non-PER physics faculty who are ``representative of likely adopters of PER-based instructional strategies"~\cite{henderson2007barriers} (pg. 2). They found that although these instructors' beliefs about teaching largely aligned with research-based practices, they faced situational barriers that hindered their adoption of these methods. These barriers often exist outside of instructors' direct control and include large class sizes, pressure to cover content, and students being accustomed to passive learning. In a review study by B\o{}rte \textit{et al.},~\cite{borte2023barriers} the authors identified three groups of barriers to active learning use: leadership and organizational structures which do not support innovative teaching methods (e.g., rewarding research more than scholarly teaching), teaching competence and lack of training of instructors, and limited technology access in active learning classrooms (e.g., immobility of learning equipment to different locations). Similar sets of barriers have been identified across other existing literature~\cite{henderson2007barriers, gardner2021exploring}. 


\subsection{Current study}

The aforementioned studies have the overarching goal of promoting and improving the uptake of active learning methods in science instruction. We aim to contribute to this same goal in the current study, however we focus on specific named active learning methods in physics (SCALE-UP, ISLE, Tutorials, and Peer Instruction) rather than research-based instructional strategies broadly. We also take a more comprehensive approach, rather than looking at relevant features of adoption and implementation in isolation. That is, we capture a holistic picture of instructors who are established users of these named active learning strategies, including their background, training they have received, their pathways to using named methods, and supports and difficulties they have experienced in their implementation (in the remainder of this paper, we use ``difficulties" instead of ``barriers" when referring to challenges faced by instructors because in our sample, challenges did not prevent implementation). Characterizing all of these aspects of named method users has a strong potential to bridge the gap between research and practice, as the full lived experiences of instructors can provide actionable insights for both method developers seeking to refine and disseminate their methods and future implementers seeking to navigate adoption in their own contexts.   


Additionally, by looking across several named methods, we aim to distinguish patterns that may be unique to specific active learning strategies, moving beyond a monolithic categorization of ``active learning." Lund and Stains~\cite{lund2015importance} found that contextual differences across scientific disciplines may affect the uptake of active learning strategies in each discipline. We hypothesize that even within the discipline of physics, different named active learning methods comprise different contexts, with varying goals and logistical requirements, that coincide with different types of dissemination pathways and supports for and difficulties in implementation. For example, SCALE-UP relies on a specific classroom layout and integrated class meetings. These method-specific requirements may necessitate more administrative support compared to other methods to build new classroom spaces and/or arrange course schedules where lecture, lab, and recitation sections are merged~\cite{foote2016enabling}. Similarly, ISLE places emphasis on scientific processes~\cite{etkina2007investigative}, often requiring a curriculum that is rooted in and structured around experimentation. Therefore, administrative support may also be important when adopting this method. On the other hand, Tutorials and Peer Instruction focus on conceptual understanding and are often used as supplements to existing course structures and curriculum rather than full course overhauls~\cite{dancy2024physics}. The types and extents of supports that these two methods require may be different than SCALE-UP and ISLE. To examine these possibilities, we determine whether and how different patterns of dissemination and/or sustained implementation emerge across the four named methods we investigate, what features of the methods might explain these differences, and the relevant implications for supporting broader use of each method. Overall, we aim to address the following research questions:

\begin{enumerate}
    \item What are the institutional contexts and instructional backgrounds (e.g., teaching experience and training) of introductory physics instructors who implement one of four named active learning methods (i.e., SCALE-UP, ISLE, Tutorials, or Peer Instruction)?
    \item When and how do introductory physics instructors first learn about named active learning methods (i.e., SCALE-UP, ISLE, Tutorials, or Peer Instruction), and what factors support or hinder their implementation of these named methods?
\end{enumerate}

\section{Methods}

The findings we report in this work are based on semi-structured interviews with introductory physics and astronomy instructors. In this section, we first describe our sample of instructors and data collection procedures and then provide details about the data analysis methods.

\subsection{Data collection}

This study is a follow up to the second iteration of a national project called Characterizing Active Learning Environments in Physics (CALEP). While the first iteration of the project (CALEP1) aimed to characterize known high-fidelity implementations of six named active learning methods~\cite{commeford2022characterizing,commeford2021characterizing}, the second iteration (CALEP2) focused on the implementations and impacts of a broad set of instructors using four named active learning methods~\cite{sundstrom2026beyond, sundstrom2026relative, bukola2025measuring, sundstrom2025evaluating, bukola2025perc}. As part of CALEP2, we recruited 31 introductory physics or astronomy instructors across the United States who self-reported using one of four named methods: SCALE-UP, ISLE, Tutorials, and Peer Instruction. We recruited the instructors through the research team's own networks, advertisements posted on the American Physical Society website, and snowball sampling. Participating instructors collected classroom observation data, pre- and post-semester network surveys, and pre- and post- semester concept inventory data in fall 2023, spring 2024, or fall 2024. 

For the current study, we aimed to recruit as many instructors as possible from the 31 instructors who had previously participated in the CALEP2 project. These instructors were best suited for our work because they were known to be named method implementers and their previous data provided means to probe specific aspects of their implementation and triangulate their responses with actual classroom practices. The last author (M.S.) initially sent out emails to all 31 instructors in spring 2026, with two follow-up emails sent out at two-week intervals to instructors who had not yet volunteered to participate, after which we stopped recruitment. The recruitment emails included an overview of the interview study, details about compensation, and a link to schedule a 1.5 hr long interview through Calendly (an online scheduling platform). In total, 20 of the 31 instructors opted to participate in our interview study, representing a 65\% response rate. 

The 20 instructors were from 17 unique institutions in the United States. Four instructors were SCALE-UP users, five were ISLE users, four were Tutorials users, and seven were Peer Instruction users. Three of the five ISLE instructors were from the same institution and two of the seven Peer Instruction instructors were from the same institution. All SCALE-UP and Tutorials instructors were from different institutions. Apart from two of the Tutorials instructors who taught introductory astronomy, all other instructors taught introductory mechanics. We provide further details about our sample of participants as part of the Results (see Section~\ref{whoarethey}).

We developed the interview protocol (see Appendix A) to address the different aspects of instructors' use of active learning methods highlighted in Section~\ref{intro}, including their teaching background, participation in training opportunities, pathways to using named methods, and supports and difficulties they faced during implementation. The protocol also included prompts that addressed instructors' use and modifications of certain components of their named method and instructors' perceptions on students questions. Most responses related to these prompts were not included in our current analysis, but they are relevant for future projects. To ensure that the interview protocol was worded clearly and elicited responses that aligned with our research questions, we conducted one pilot interview with an instructor from the research team's professional network (who did not contribute data to CALEP2) who identified as a mixed user of SCALE-UP (i.e., in addition to SCALE-UP elements, they used components of other named methods). Based on this interviewee's responses and feedback, we made minor modifications to the wording of some of the questions in the protocol. Data from the pilot interview were not included in our analysis.

Following the practices recommended by McPadden \textit{et al.}~\cite{mcpadden2023planning} to improve participant accessibility in qualitative research studies, we sent participants the full interview protocol 48 hours before their scheduled session and asked whether they required any accommodations. One instructor requested to split their interview into shorter sessions; for this participant, we conducted two 30-minute interviews within the same week rather than a single, longer session. All other interviews took place during a single session. With all 20 participants, we conducted semi-structured interviews via Zoom (a cloud-based meeting software). The semi-structured format allowed the interviewers to ask follow-up questions beyond the initial interview protocol, which helped to uncover more depth from instructors' responses and ultimately provided richer data to address our research questions~\cite{ruslin2022semi}. 

All interviews were conducted in the spring of 2026. Interview durations ranged from 31 minutes to 1 hour and 28 minutes (all questions on the protocol were asked to every participant) and were video-recorded with the consent of participants. The first and last authors (I.B. and M.S.) were present for all interviews. I.B. read the interview questions aloud and asked most of the follow-up questions based on the interviewee responses. M.S. posted the questions in the Zoom chat window, providing an alternative way for instructors to access and recall the questions, and also asked follow-up questions in some cases.  Each participant received a \$100 Amazon gift card as compensation for their time.


\begin{figure*}[t]
\includegraphics[width=1\textwidth]{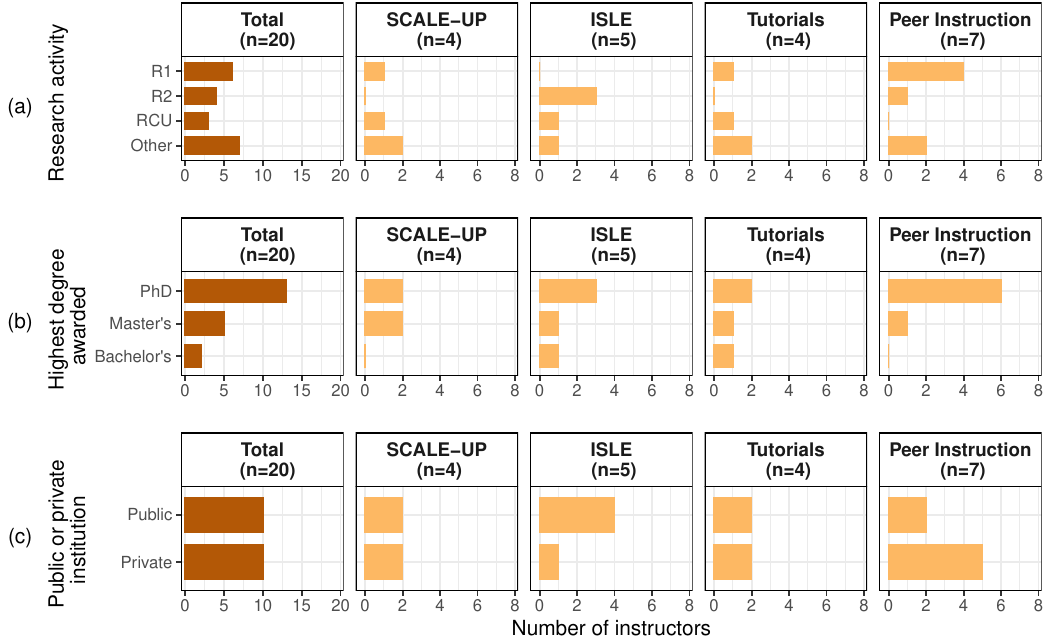}
    \caption{Number of instructors (total and by method) at different institution types, based on the 2026 Carnegie Classifications. We categorized institutions by (a) research activity (R1 indicates ``Very High Research Spending and Doctorate Production," R2 indicates ``High Research Spending and Doctorate Production," and RCU indicates ``Research Colleges and Universities"), (b) highest degree awarded, and (c) public or private institution.
    }\label{instiution-info}
\end{figure*}

\subsection{Data analysis}

All 20 interviews were automatically transcribed by Zoom. I.B. manually checked and modified all transcripts by listening to the Zoom recordings, making spelling and grammar edits, and noting non-verbal cues that were relevant for interpreting interviewees' responses (e.g., air quotes). I.B. also re-formatted the transcripts from the auto-generated Zoom format to blocks of interviewer questions and interviewee responses to facilitate analysis. Being immersed in the data allowed I.B. to gain familiarity with interviewee responses and note any first impressions~\cite{tesch2013qualitative}. 

To organize the data for analysis, I.B. identified all parts of each transcript (i.e., including interviewee responses to all questions) that were relevant for aspects of instructors' experiences we aimed to analyze and compiled them into one spreadsheet. These aspects included experience with Discipline-Based Education Research (DBER), training received on Evidence-Based Instructional Practices (EBIPs) in general and named methods specifically, number of years spent teaching, means of initial exposure to their named method, their reason for using their method, and supports for and difficulties in implementation of the method. The spreadsheet contained one row per participant and one column per aspect. In one case where the instructor's response about their teaching roles and number of years they spent in those roles was unclear, I.B. contacted the instructor by email to provide clarification and included this information in the spreadsheet. 

For each aspect of instructors' experiences, we conducted an inductive qualitative content analysis~\cite{drever1995using}. First, I.B. and M.S. read through all of the relevant data for the topic. Then, I.B. identified distinct ideas within the data, organized them into broader themes, and drafted an emergent coding scheme. Next, I.B. and M.S. engaged in negotiated discussions to iteratively modify and reach a consensus coding scheme, and then applied the coding scheme to each participant's responses together~\cite{campbell2013coding}. We did not calculate IRR because of the small sample size in this study and because the two coders discussed coding until full consensus was reached. Some coding schemes were mutually exclusive, while others were not. We note these different cases throughout the Results section. 

In addition to developing and applying these emergent coding schemes, we categorized instructors' current institutions using the 2026 Carnegie Classifications available online. These were the only data included in our findings that were not directly prompted in the interviews. 

We then created bar charts to visualize the frequency of codes and the institutional classifications, both aggregated and for each named active learning method. This approach provided a holistic yet descriptive picture of the backgrounds of instructors who adopt named active learning methods in physics and their experiences in implementing the methods. We noted the relative frequencies of different codes across all instructors and for instructors using each named method, and used quotes from the interview transcripts to substantiate any interpretations.  Our analysis should be considered exploratory, as we did not aim to quantitatively compare or make any statistical claims about code frequencies across different groups of instructors.

\section{Results}

Here, we present our findings by research question. For each research question, we first summarize the overarching findings and then describe more nuanced patterns and specific examples across the four named active learning methods. 
For all example quotes, ellipses indicate retracted portions of the responses for the sake of brevity, bracketed text indicates that the research team inserted or changed text to provide necessary context or fix grammatical mistakes, and institution and individual names are removed to preserve anonymity. We also distinguish which instructor each quote came from by labeling it with the named method and a number (e.g., SCALE-UP 1). In this section's figures, orange bars indicate that codes were mutually exclusive, while blue bars indicate that codes were not mutually exclusive.

\begin{table*}
\centering
\caption{Emergent codes, including definitions and example quotes, related to instructors' engagement with DBER. ``DBER researcher" was coded exclusively, while interviewees who were not coded as DBER researchers could be assigned multiple of the three other emergent codes. We assume that the instructors who are DBER researchers participate in the other three forms of engagement. \label{DBEREngagement-table}}
\begin{ruledtabular}
\begin{tabular} {>{\hangindent=5mm\hangafter=1}p{0.2\linewidth} >{\hangindent=1em\hangafter=1} p{0.3\linewidth} >{\hangindent=1em\hangafter=1}p{0.46\linewidth} }

Code  & Definition & Example (Method)\\
   \hline
    DBER researcher & &\\
     & Previously conducted and/or actively conducts DBER, including publishing, mentoring, or collaborating on DBER projects & \textit{``I did some physics education research as an undergraduate, and then that was my graduate school, and then that's been my, like, professional research career. So that's been, like, almost 15 years at this point, in one way or another."} (Tutorials 3)\\ 
    
    Non-DBER researcher & &\\
     & &\\
     \hspace{5mm}Engaged with DBER literature & Reads DBER papers or articles and keeps up with journals (e.g., \textit{The Physics Teacher}, \textit{American Journal of Physics}, \textit{Physical Review Physics Education Research}) & \textit{``I read Physics Today. I'm always interested in whatever physics education or astronomy education article they have in Physics Today... I'll see them come across my newsfeed. I'll click on things that I find interesting."} (Tutorials 1)\\
     
     
    & &\\
    \hspace{5mm}Engaged in structured DBER activities & Attends DBER conferences, workshops, and/or seminars; part of DBER-related associations (e.g., Professional development for Emerging Education Researchers (PEER), Advanced Laboratory Physics Association (ALPhA)) & \textit{``Recently, I've actually been part of PEER... So I actively participated in one of their, like, week-long workshops last summer because I wanted to learn how to, like, get into education research and discipline-based education research in particular."} (ISLE 2)\\
     & &\\
    \hspace{5mm}Engaged in informal DBER activities & Receives information about DBER through informal channels, such as interactions with colleagues and/or DBER researchers they know & \textit{``Chatting with colleagues here, for example, we have now a physics education research faculty member... and associating with [them] and other people that have been involved with physics education research, I've gained some understanding and experience with the process... and why we feel that discipline, or evidence-based learning... is appropriate."} (Peer Instruction 3)\\
    
\end{tabular}
\end{ruledtabular}
\end{table*}

\begin{figure*}[ht]
\includegraphics[width=0.98\textwidth]{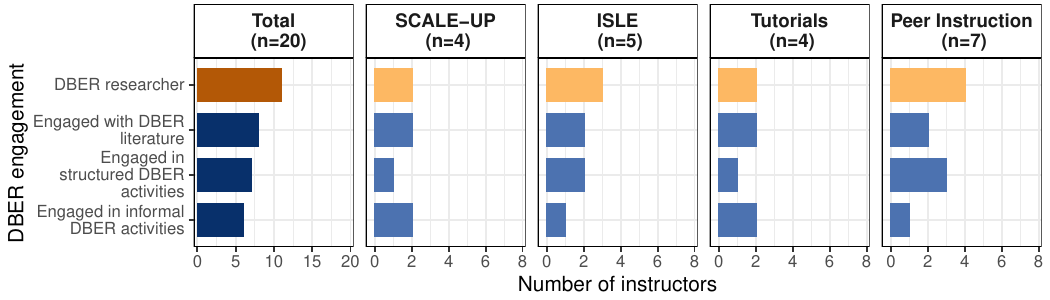}
    \caption{Frequency of instructors (total and by method) whose responses indicate the presence of emergent codes related to engagement in DBER (Table~\ref{DBEREngagement-table}). ``DBER researcher" was coded exclusively, while interviewees who were not coded as DBER researchers could be assigned multiple of the three other emergent codes. 
    }
    \label{DBERengagement-plot}
\end{figure*}

\subsection{Who are the instructors using named active learning methods?}
\label{whoarethey}

Overall, our sample of instructors implemented named methods at different types of institutions in the United States, spanning a range of research- and teaching-focused institutions and public and private institutions. The instructors were all engaged in DBER, albeit to varying degrees, and they all received some form of training on both EBIPs generally and their named method specifically. Method-specific training varied across active learning methods, with fewer SCALE-UP users reporting engagement in formal professional development than users of the other three methods and most ISLE users reporting multiple forms of method-specific training.

\begin{table*}[ht]
\centering
\caption{Emergent codes, including definitions and example quotes, related to instructors' training on EBIPs. Codes were not mutually exclusive. \label{Training-EBIP-table}}
\begin{ruledtabular}
\begin{tabular} {p{0.15\linewidth} >{\hangindent=1em\hangafter=1} p{0.35\linewidth} >{\hangindent=1em\hangafter=1}p{0.46\linewidth} }

     Code  & Definition & Example (Method)\\
   \hline
    Low intensity professional development & Attended brief and/or one-off training programs about EBIPs, such as Teaching Assistant (TA) training as a graduate student, one-time workshops, relevant undergraduate or graduate courses, and conferences talks &  \textit{``Over these years, I guess I did attend workshops? [For] example, the latest one was at the January AAS [American Astronomical Society] meeting. They had a teaching workshop. It's [an] astronomy national meeting, so I attended that."} (Peer Instruction 7) \\
    & &\\
    High intensity professional development & Participated in extensive, structured training programs about EBIPs, such as week- or semester-long workshops, institutes (e.g., the American Physical Society Faculty Training Institute), and fellowship programs & \textit{``There's a thing called the Institute of Scientists and Engineer Educators, ISEE. They held a multi-stage workshop. It went from, like, a conference in California to a meeting somewhere in Maryland. And this was 100 hours of training on inquiry and equity-based teaching practices."} (ISLE 1)\\
  & &\\
    Post-secondary education & Received a Bachelor's (major or minor), Master's, or doctoral degree in science education or physics education research (PER) & \textit{``I did take all of the required, like, education courses as an undergraduate student, where my... bachelor's degree is [in] physics education. So, I think that kind of corresponds to, like, a minor in education."} (Peer Instruction 4)\\  
    & &\\
    Informal professional development & Received training about EBIPs through less structured means, such as through mentorship from faculty at their institution or reading groups & \textit{``I've been part of some reading groups on some evidence-based pedagogy papers, both, like, in physics and, like, cross-disciplinary, at like, scattered times over the past 7 or 8 years."} 
     (SCALE-UP 2)\\
    
    \end{tabular}
\end{ruledtabular}
\end{table*}

\begin{table*}[ht]
\centering
\caption{Emergent codes, including definitions and example quotes, related to instructors' training for their named method. Codes were not mutually exclusive. \label{Training-method-table}}
\begin{ruledtabular}
\begin{tabular} {p{0.14\linewidth} >{\hangindent=1em\hangafter=1} p{0.35\linewidth} >{\hangindent=1em\hangafter=1}p{0.46\linewidth} }

     Code   & Definition & Example (Method)\\
   \hline

     Formal professional development & Participated in or attended formal trainings about the method, such as specific TA training as a graduate student, workshops, seminars, talks, conferences, or courses & \textit{``The course that I took... reading these papers [on] Peer Instruction when I first got to [institution]... was definitely, like, formal instruction on how to do this... We had, like, homework, we were supposed to, like, read the paper, and then we came in, and then we, like, talked about it and, you know, did these kinds of active learning things... about how to do them, right? So, like, [in] the course itself, we use[d] those things in class to learn the stuff."} (Peer Instruction 4)\\
     
     & &\\
    Direct engagement & Observed and/or worked with method developers or instructors using the method, for example visiting sites where the named method is implemented, working as a graduate TA using the method, or training others on how to use method & \textit{``When I was a TA at [institution], I think I did at least one semester [as the lead] instructor for a studio section of the course."} (SCALE-UP 1)\\

     & &\\
    Informal professional development & Learned about the method through reading literature and/or interacting with colleagues who implement the method & \textit{``Since I start[ed] to work with [PER researcher who uses ISLE], I have been informally trained by [them], and when I was in [institution], I frequently called [them] and [asked them] questions about ISLE and how to implement it in [institution] with certain class size, format, right? [They] gave me a lot of support."} (ISLE 5)\\
    
\end{tabular}
\end{ruledtabular}
\end{table*}

\begin{figure*}[!htbp]
\includegraphics[width=0.98\textwidth]{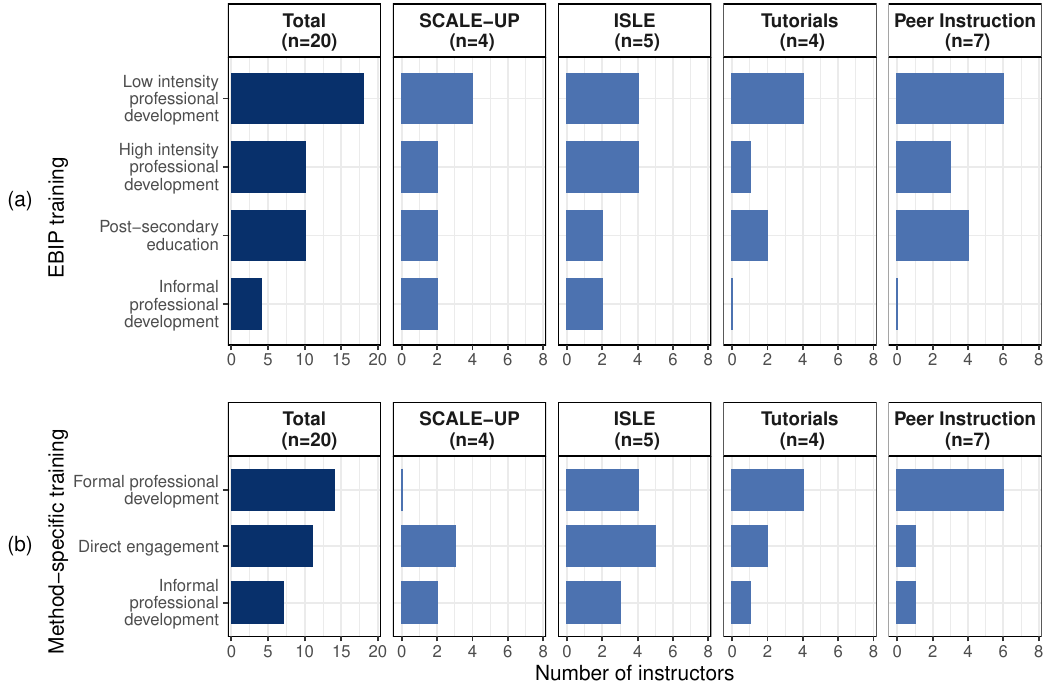}
    \caption{Frequency of instructors (total and by method) whose responses indicate the presence of emergent codes related to their training experiences with (a) EBIPs and (b) their named method (Tables~\ref{Training-EBIP-table} and~\ref{Training-method-table}).}\label{Training-background-plot}
\end{figure*}
\subsubsection{Instructional context}

To gain an understanding of where the physics instructors in our study were implementing named active learning methods, we categorized their institution based on research activity, highest degree awarded, and whether the institution is public or private (Fig.~\ref{instiution-info}). The instructors were evenly split between research-focused institutions (R1 and R2) and primarily teaching institutions (RCU and Other; first panel of Fig.~\ref{instiution-info}a), as well as between public and private institutions (first panel of Fig.~\ref{instiution-info}c). More than half of the instructors were at PhD-granting institutions, with fewer instructors at Master's- or Bachelor's-granting institutions (first panel of Fig.~\ref{instiution-info}b).

There was some variation in these distributions of institution types across the four named methods. Most of the SCALE-UP and Tutorials instructors in our sample were at teaching-focused institutions (second and fourth panels of Fig.~\ref{instiution-info}a), while most ISLE and Peer Instruction instructors were at research-focused institutions (third and fifth panels of Fig.~\ref{instiution-info}a). Instructors using SCALE-UP and Peer Instruction were exclusively at PhD- or Master's-granting institutions (second and fifth panels of Fig.~\ref{instiution-info}b), though the majority of Peer Instruction users were at PhD-granting institutions. This pattern may be due to graduate degree-granting universities typically having larger undergraduate class sizes, for which SCALE-UP and Peer Instruction were explicitly designed. We note again that three of the five interviewed ISLE instructors were from the same institution, as well as two of the seven Peer Instruction users.


\subsubsection{Engagement with DBER}

All of the instructors in our study were either directly part of or engaged with the DBER community (Table~\ref{DBEREngagement-table} and Fig.~\ref{DBERengagement-plot}). 11 of the 20 interviewed instructors were DBER researchers (first panel of Fig.~\ref{DBERengagement-plot}). For the instructors who did not identify as DBER researchers, engagement was fairly evenly distributed across other modes, including engaging with DBER literature, structured DBER activities, and informal DBER activities (first panel of Fig.~\ref{DBERengagement-plot}). Eight of the nine non-DBER instructors reported more than one form of engagement with the field. These patterns also held \textit{within} each of the four named methods: about half of the instructors using each method were DBER researchers, and the remaining non-DBER researchers reported engagement spread across the other three activities (second to fifth panels of Fig.~\ref{DBERengagement-plot}).

\subsubsection{Training}

We asked instructors about the types of training they had received, if any, for EBIPs in general and for their named method specifically. Instructor responses related to each type of training yielded different sets of emerging codes (Tables~\ref{Training-EBIP-table} and~\ref{Training-method-table}). All 20 instructors reported receiving training on both EBIPs and their named method, which included different extents (low intensity or high intensity) and types (formal or informal) of professional development and graduate-level coursework and/or degrees in education (first panel of Figs.~\ref{Training-background-plot}a and~\ref{Training-background-plot}b). 16 out of the 20 instructors indicated receiving more than one form of training for EBIPs.


 \begin{figure*}[!htbp]
\includegraphics[width=0.98\textwidth]{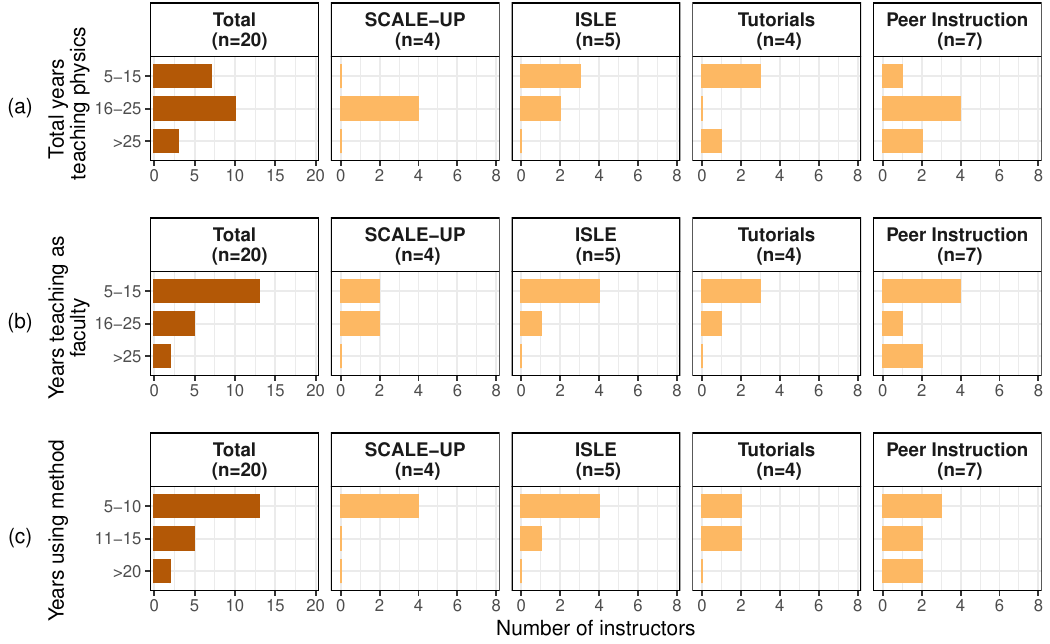}
    \caption{Number of instructors (total and by method) with different teaching experience as measured by (a) total teaching years (in all physics teaching roles), (b) years as faculty (including time as visiting professors, adjunct faculty, tenure track or other professor related roles), and (c) years using named method (no instructor fell within the 16-20 years bin). 
}\label{TeachingYears-plot}
\end{figure*}

We observed some variation in training across the named methods, especially for method-specific training. None of the SCALE-UP instructors received training through formal professional development avenues, such as workshops or conferences (second panel of Fig.~\ref{Training-background-plot}b). 
Instead, the SCALE-UP instructors identified informal professional development opportunities and direct engagement with the method (e.g., observing the method in practice and/or implementing the method themselves under the supervision of a faculty member) as their primary training experience. One SCALE-UP instructor (SCALE-UP 1), for example, indicated receiving direct engagement training through teaching with the method as a TA: \textit{``When I was a TA at [institution], I think I did at least one semester [as the lead] instructor for a studio section of the course"} (second row of Table~\ref{Training-method-table}). Instructors also flagged that SCALE-UP may not have as many formal professional development avenues as other methods; for instance, one instructor (SCALE-UP 4) mentioned that \textit{``there was no formal training on SCALE-UP,"} which was \textit{``one of the difficulties"} they faced in using the method. 

All ISLE instructors reported receiving method-specific training through direct engagement with the method and they all reported receiving more than one form of method-specific training (third panel of Fig.~\ref{Training-background-plot}b). 
Most of the interviewed Tutorials and Peer Instruction instructors received method-specific training through formal professional development (fourth and fifth panel of Fig.~\ref{Training-background-plot}b). For example, one Peer Instruction user (Peer Instruction 4) said, ``\textit{The course that I took... reading these papers [on] Peer Instruction when I first got to [institution]... was definitely, like, formal instruction on how to do this... We had, like, homework, we were supposed to, like, read the paper, and then we came in, and then we, like, talked about it and, you know, did these kinds of active learning things... about how to do them, right? So, like, [in] the course itself, we use[d] those things in class to learn the stuff}" (first row of Table~\ref{Training-method-table}). 


\subsection{How do instructors learn about and sustain implementation of named methods?}
Instructors learned about named active learning methods through different avenues and began using them at different timepoints in their teaching trajectory. There were also a variety of supports for implementing named methods, including from community members, administrative bodies, and instructors themselves. On the other hand, instructors faced a range of difficulties, some of which are reminiscent of long-standing issues associated with using active learning and some of which are nascent issues in the contemporary science classroom.

\begin{table*}[t]
\centering
\caption{Emergent codes, including definitions and example quotes, capturing how instructors first learned about the named method they use. Codes were mutually exclusive.
\label{MethodHistory(a)-table}}
\begin{ruledtabular}
\begin{tabular} {p{0.11\linewidth} >{\hangindent=1em\hangafter=1} p{0.35\linewidth} >{\hangindent=1em\hangafter=1}p{0.46\linewidth} }

   Code  & Definition & Example (Method)\\
   \hline
    Direct engagement & Had direct interaction with the method, for example, as a TA for an instructor using a similar approach, being a student in a course where the instructor used the method, or being a graduate student of an advisor who designs and/or conducts research on the method &  \textit{``That would be when I became... [method developer]'s PhD student."} (ISLE 4)\\
     & &\\
    Professional development & Learned about the method through conferences, workshops, coursework, institutional resources for instructors, and/or other trainings & \textit{``I mean... it would have been when I started graduate school at [institution]. They give this workshop for new teaching assistants, and that workshop includes information about Peer Instruction."} (Peer Instruction 5) \\ 
    & &\\
    Informal avenues & Learned about the method through informal channels, such as an email list or conversations with faculty & \textit{``Through this Astro Learner email list, and [method developer] is on the Astro Learner group, and so every once in a while, [they'd] be like, `Oh, we're doing a CAPER [Center for Astronomy and Physics Education Research] Tutorial workshop, or we just put out some new CAPER stuff.' And, so I was like, `Oh, that's interesting.' So I got a workbook."} (Tutorials 1)\\
     & &\\
    Institutional decision & Learned about the method because it was being used at their institution & \textit{``Well, you know, actually, pretty early on, when I got to [institution], because I think even then, they were doing cooperative learning in the discussion classes... So probably, yeah, right after I got to [institution], they were doing cooperative learning."} (Peer Instruction 2)\\
    
\end{tabular}
\end{ruledtabular}
\end{table*}

\begin{table*}[ht]
\centering
\caption{Emergent codes, including definitions and example quotes, capturing instructor's reasons for using their named method. Codes were not mutually exclusive.
\label{MethodHistory(b)-table}}
\begin{ruledtabular}
\begin{tabular} {p{0.1\linewidth} >{\hangindent=1em\hangafter=1} p{0.35\linewidth} >{\hangindent=1em\hangafter=1}p{0.5\linewidth} }

   Code  & Definition & Example (Method)\\
   \hline
    Instructor values & Aspects of the method reflect the instructor's personal values, such as promoting student interaction and following best teaching practices & \textit{``Once I became aware of... how little students benefit from passively watching lecturing... I wanted to switch to active learning... There's no other way, I think, to have students truly grapple with the material without them answering questions about the material... The Tutorials point very strongly to the areas where the students need work."} (Tutorials 4) \\ 
    & &\\
    Institutional factors & Aspects of the method align with institutional structures (e.g., class sizes, modality), institutional values (e.g., active learning), and/or institutional decisions (i.e., the institution or department decided to implement the method) &  \textit{``For teaching, the idea is, like, to do as much active learning as possible... within the constraints of the environment that we have... Right now, I'm teaching a lecture with 224 students, and... some of the things that we might do in a smaller class are not really possible in that setting, but Peer Instruction is something that can be scaled... to that size of class."} (Peer Instruction 1)\\
     & &\\
    Student outcomes & Aspects of the method support student outcomes, for example dealing with common misconceptions, improving student learning or peer recognition, and enhancing scientific practices &  \textit{``I want my students to, like, actively be part of, like, the discovery of these... physical laws, and always feel capable of, like, testing any of the things that I'm saying. Like, they shouldn't just take [it] from me... that, like, Newton's first law is true. It's like, we should have, like, some observations, and, like, really test... this theory. So I really loved those aspects of ISLE."} (ISLE 2)\\
     & &\\
    Ease of implementation & Aspects of the method make it easy to adopt, such as flexibility in the way it is used, clear underlying philosophy, already-aligned curriculum (i.e., with course learning objectives), and availability of resources; instructor has previous experience with or training for the method to facilitate use & \textit{``SCALE-UP, the way I would put it, is the most flexible environment for doing different things. If you have the space, if you have a classroom that's set up as a SCALE-UP, you can decide one day I'm gonna do worksheets, and decide another day I'm gonna do clicker questions, and decide one day I'm gonna do an interactive lab demonstration. And you could do it all seamlessly. And that sort of flexibility is really what I... like about it."} (SCALE-UP 3)\\
    
\end{tabular}
\end{ruledtabular}
\end{table*}

\begin{figure*}[ht]
\includegraphics[width=0.98\textwidth]{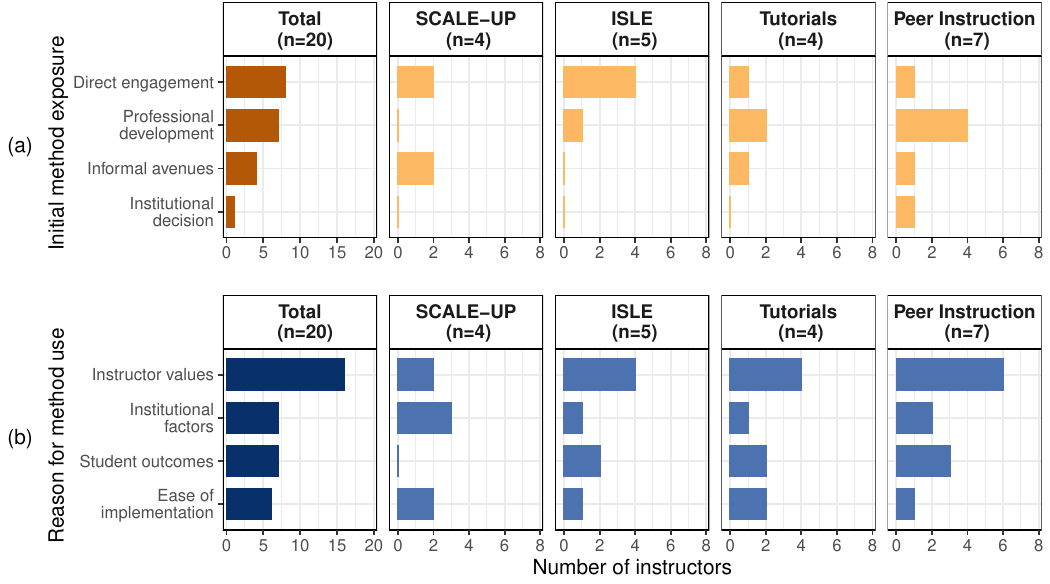}
    \caption{Frequency of instructors (total and by method) whose responses indicate the presence of emergent codes related to (a) how they first learn about the named method they use and (b) the reason(s) they decided to use their named method. Definitions of the codes and example quotes can be found in Tables~\ref{MethodHistory(a)-table} and~\ref{MethodHistory(b)-table}.
}\label{MethodHistory-plot}
\end{figure*}

\subsubsection{Pathways to using named methods}

Instructors' previous teaching roles and experience were closely linked to the pathway through which they came to use their named method. While most of the instructors had been using their named method for 5 to 15 years (first panel of Fig.~\ref{TeachingYears-plot}c) at the time of the interview, many had been teaching physics for much longer than that (as high school teachers, graduate students, and/or faculty members; first panel of Fig.~\ref{TeachingYears-plot}a). This means that most instructors did not use their named method throughout their entire teaching career; rather, they used a different teaching style and later learned about the active learning method. One Tutorials instructor (Tutorials 4) stated that it was \textit{``unfortunate"} that they had mostly used passive lecturing as an instructor in the past, but that they were \textit{``unaware of, like, clickers and stuff at that time."} In some cases, instructors previously used their named method, then did not use the named method for a period of time due to the nature of their role (e.g., as a graduate TA for a different course or as a visiting professor), and then resumed use. 


Relatedly, there are more instructors who had used their named method for 5-15 years than instructors who had been teaching as faculty for 5-15 years (first panel of Figs.~\ref{TeachingYears-plot}b and~\ref{TeachingYears-plot}c), indicating that some instructors were using their method before becoming faculty, for example as a graduate TA. When asked how long they had been using the method, one Tutorials instructor (Tutorials 3) who had used the method as a graduate student and was now using it as a faculty member stated, \textit{``the whole time. But yeah, it was, like, my decision for the past, like, five years."} Similarly, one Peer Instruction instructor (Peer Instruction 6) noted that they \textit{``came right into [their] teaching assistant gig with instructors who wanted to apply it."}

Interestingly, the only instructors who had used their named method for longer than 20 years were Peer Instruction users (fifth panel of Fig.~\ref{TeachingYears-plot}c). Of the four methods included in this study, Peer Instruction was the first to have been published and widely circulated within the physics education community~\cite{mazur1997peerconf}. SCALE-UP dissemination began shortly after~\cite{beichner1999student}, though all of the SCALE-UP instructors in this study started using the method within the past 5 to 10 years (second panel of Fig.~\ref{TeachingYears-plot}c). 


\begin{table*}[ht]
\centering
\caption{Emergent codes, including definitions and example quotes, related to the supports instructors received for implementing their named method. Codes were not mutually exclusive.
\label{MethodSupport-table}}
\begin{ruledtabular}
\begin{tabular} {p{0.14\linewidth} >{\hangindent=1em\hangafter=1} p{0.32\linewidth} >{\hangindent=1em\hangafter=1}p{0.47\linewidth} }

   Code  & Definition & Example (Method)\\
   \hline
    Community support & Received operational support from DBER groups, method experts, and/or faculty colleagues when implementing the method & \textit{``PER as a whole community supported me, for sure. Yeah, the teaching environment I'm in supported me, allow[ing] me... to fail, to make mistakes, and make adjustments. So, of course, the ISLE people [who gave] me an opportunity to participate in facilitating the workshop that strengthened... my understanding of how to implement ISLE."} (ISLE 5)\\ 
    & &\\
    Instructor agency & Departments give instructors freedom to decide how they teach and/or encourage active learning &  \textit{``I basically had free reign. Like, I could do what I wanted with the physics curriculum, teach it however I wanted, so that flexibility was wonderful. I didn't have... anyone telling me, like, here's what we did before, here's how you're supposed to do it. So that was... pretty great. I really could just follow what I thought would be best."} (ISLE 2)\\
     & &\\
    Administrative support & Received operational support from their institution and/or department to use the method, for example through technological help, access to peer facilitators, use of rooms set up for active learning, or time off for instructors to develop course material & \textit{``Well, clearly the work that [faculty name] and [faculty name] did to revamp the class is the main thing, I think, right? Because the physics department gave them, I think, an entire year off teaching to re-work this whole class, so that was a lot of support right there."} (Peer Instruction 2)\\
     & &\\
    Ease of implementation & Method is easy to implement, for example because there are existing resources to help create course material and free access to necessary software & \textit{``Like the ease of implementing all of that. I mean, just having that website is fantastic."} (Tutorials 2)\\
     & &\\
    Instructor background & Felt comfortable using active learning, received training on active learning, conducted education research, or has experience using the method & \textit{``Certainly, my preparation has been the biggest support... And that carried me through a lot, including being able to be a super confident person when the architect talked to me, or when they were discussing building the room, saying, these are the things I need, and then coming back and saying, well, can you live with this? And me saying yes or no. Like, if I hadn't been through what I had been through, I wouldn't know how to answer those questions, so that's the biggest thing."} (SCALE-UP 3)\\
     & &\\
    Feedback & Received positive and/or constructive feedback from students and/or faculty & \textit{``Anytime I have faculty members come and, like, observe my class, they're... really happy with that. And I get positive feedback from that, and I get really positive feedback from students... Something as simple as that is... it's very rewarding, right?"} (Peer Instruction 6)\\
    
\end{tabular}
\end{ruledtabular}
\end{table*}

During these teaching journeys, instructors first learned about their named method in a variety of ways (Table~\ref{MethodHistory(a)-table} and Fig.~\ref{MethodHistory-plot}a). Direct engagement with the method (e.g., serving as a TA for a course using the method or working with an advisor who uses or conducts research on the method as a graduate student) and professional development avenues (e.g., workshops and conferences) were the most common ways the instructors in our sample became aware of their named method (first panel of Fig.~\ref{MethodHistory-plot}a). However, initial exposure varied across the named methods. SCALE-UP instructors learned about the method either through direct engagement with the method, for example by being a TA for a SCALE-UP class, or through informal avenues, such as conversations with faculty using the method (second panel of Fig.~\ref{MethodHistory-plot}a). 
Notably, none of the SCALE-UP instructors learned about the method through professional development, in line with our earlier finding that none of the SCALE-UP instructors received method-specific training through formal professional development (Fig.~\ref{Training-background-plot}b). 
ISLE instructors primarily learned about the method through direct engagement, as many of them were trained by other ISLE users and/or were involved in training future users (third panel of Fig.~\ref{MethodHistory-plot}a). When asked how they knew about the method, one ISLE instructor (ISLE 4) responded, \textit{``That would be when I became... [method developer]'s PhD student"} (first row of Table~\ref{MethodHistory(a)-table}), indicating that their knowledge of the method came from working directly with the method developer. Compared to the other methods, Tutorials and Peer Instruction were more often diffused through professional development avenues (fourth and fifth panels of Fig.~\ref{MethodHistory-plot}a), in line with our previous finding that most instructors of these two methods received method-specific training through professional development (Fig.~\ref{Training-background-plot}b). One Peer Instruction user (Peer Instruction 5) described, \textit{``I mean... it would have been when I started graduate school at [institution]. They give this workshop for new teaching assistants, and that workshop includes information about Peer Instruction"} (second row of Table~\ref{MethodHistory(a)-table}).

Beyond learning that the method exists, instructors described a range of reasons for why they ultimately decided to implement it (Table~\ref{MethodHistory(b)-table}). In addition to using the method because it improved student outcomes and was easy to implement, instructor values seemed to play a crucial role in this decision across all four methods (first panel of Fig.~\ref{MethodHistory-plot}b). That is, many of the instructors mentioned that their personal beliefs about the strengths of interactive engagement drove them to use the method. For instance, an instructor using Tutorials (Tutorials 4) stated, \textit{``Once I became aware of... how little students benefit from passively watching lecturing... I wanted to switch to active learning... There's no other way, I think, to have students truly grapple with the material without them answering questions about the material... The Tutorials point very strongly to the areas where the students need work"} (first row of Table~\ref{MethodHistory(b)-table}). 
SCALE-UP is the only method for which instructor values were not the most salient reason for method use, as three out of the four SCALE-UP instructors highlighted institutional factors (second panel of Fig.~\ref{MethodHistory-plot}b). 
In these cases, instructors stated that the method resonated well with the student population or class structures at their institution, or that their department or institution had already decided to use the method. One SCALE-UP instructor (SCALE-UP 4) described, \textit{``the choice was... thrust upon me by another faculty member who [was] very excited about the method... [they] managed to convince the administration to pony up a bunch of money... to renovate the space."} 

\begin{table*}[t]
\centering
\caption{Emergent codes, including definitions and example quotes, related to the difficulties instructors face when implementing their named method. Codes were not mutually exclusive.
\label{MethodBarriers-table}}
\begin{ruledtabular}
\begin{tabular} {p{0.15\linewidth} >{\hangindent=1em\hangafter=1} p{0.35\linewidth} >{\hangindent=1em\hangafter=1}p{0.45\linewidth} }

   Code  & Definition & Example (Method)\\
   \hline
    Structural challenges & Challenges related to time or physical constraints, such as limited time to train TAs, physical classroom space, small class sizes, cutting down the number of topics covered to use active learning, or coordinating multi-instructor courses & \textit{``It's hard not having... TAs that--I mean I could spend a lot of time training them, but it's just, like, we're only allowed to work with TAs for so many hours, and so I have to be really careful about the amount of [time] I spend [on] some things versus others."} (Tutorials 2)\\ 
   & &\\
    Student engagement and learning challenges & Student-related challenges with the method, such as resistance or lack of buy-in to active learning, lack of engagement in the material or class activities, use of AI, or varying student needs in the classroom &  \textit{``The fact that we are fighting against upwards of 12 years of ingrained bad habits in students about what education is supposed to look like... they have been conditioned to think that class time is where they shut up and listen to me talk, and just write down everything I say and make sense of it later, or cry. And a lot of what we do goes against the grain of what students are used to."} (ISLE 4)\\
   & &\\
    Challenges with using resources & Difficulties with accessing or using available resources for the method, such as software issues for clicker, high textbook costs, or having a different course or student population than curricular materials (e.g., calculus-based or not, mechanics or electromagnetism, high school versus college) &  \textit{``And I had to, on Monday, be like, but I have to calculus-ify this, which just meant, like, it wasn't organically put in. It would just be, `Okay, here we're talking about slopes, but also, let's remember, if I give you a function, you can take the derivative to get the slope.' So [the calculus] always felt tacked on. It still feels kind of tacked on to me."} (ISLE 1)\\
    & &\\
    Administrative challenges & Operational challenges within the department or institution, such as insufficient LAs, lack of administrative know-how, or lack of funding for necessary equipment or personnel & \textit{``I know everything I just said is, like, with the peer facilitators, is, like, constantly at threat of, like, not having enough funding. So I think constantly having to battle to have that program funded is definitely... a detraction, for sure."} (SCALE-UP 2)\\
   & &\\
    Lack of training & Training or preparation for using the method does not exist or instructor did not use it & \textit{``Can't think of anything else but my own preparedness, I guess, to do this. Again, the old story goes, you do a PhD in physics... and you're expected to teach. You'll teach on some level, maybe it's not a regular classroom, you're mentoring or advising, and you had no training in that. Formal training, I mean, yes, I was trained, and I sat in classes over the years, if that counts, but that's not exactly the same as doing it."} (Peer Instruction 7)\\
    
\end{tabular}
\end{ruledtabular}
\end{table*}

\subsubsection{Supports for implementation}

\begin{figure*}[ht]
\includegraphics[width=0.98\textwidth]{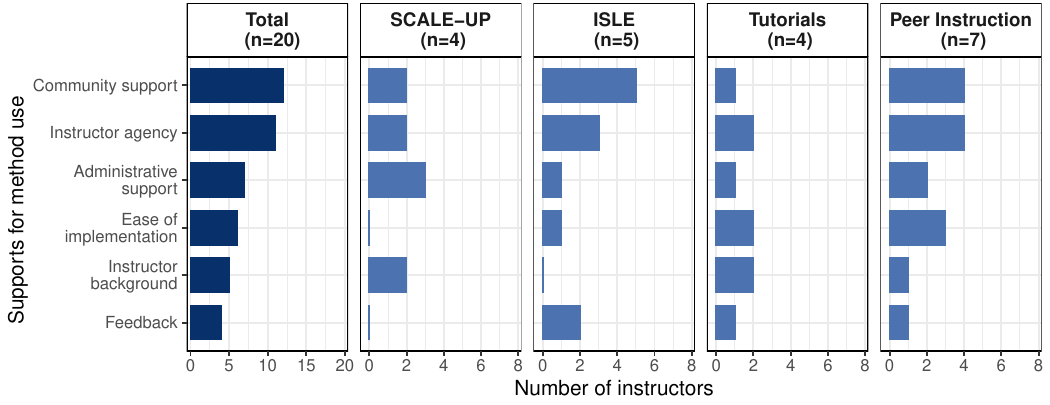}
    \caption{Frequency of instructors (total and by method) whose responses indicate the presence of emergent codes related to support they received for implementing their named methods. Definitions of the codes and example quotes can be found in Table~\ref{MethodSupport-table}.
}\label{MethodSupport-plot}
\end{figure*}

All instructors reported receiving at least one form of support for implementing their named method, and many instructors indicated receiving multiple forms of support (Table~\ref{MethodSupport-table} and Fig.~\ref{MethodSupport-plot}). Some support came from groups of people, such as the instructional community (e.g., DBER groups and method experts) and the instructor's department (e.g., through administrative support or giving instructors freedom in how they teach), while others stemmed from the nature of the method (e.g., perceived as easy to use) or from the instructor's own background and training. Community support and instructor agency to choose how they teach their physics courses were the most commonly reported forms of support across all methods (first panel of Fig.~\ref{MethodSupport-plot}), though there was some variation across individual methods. Notably, instructor characteristics (i.e., values and agency) emerged as both a reason for method use (see previous sub-section) and a form of support for instructors implementing named active learning methods.

\begin{figure*}[ht]
\includegraphics[width=0.98\textwidth]{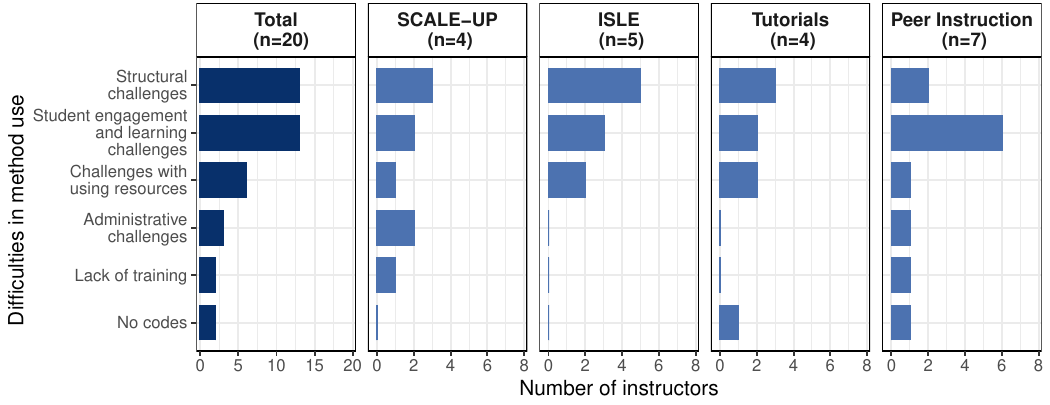}
    \caption{Frequency of instructors (total and by method) whose responses indicate the presence of emergent codes related to difficulties in method implementation. ``No codes" was assigned to instructors who reported no difficulties in implementation. Definitions of the codes and example quotes can be found in Table~\ref{MethodBarriers-table}.
}\label{MethodBarriers-plot}
\end{figure*}

Three of the four SCALE-UP instructors acknowledged administrative support, predominantly in the context of transforming their physical classroom (second panel of Fig.~\ref{MethodSupport-plot}). One user (SCALE-UP 1) described, \textit{``I mentioned the administrative aspect... Hav[ing] this kind of classroom, hav[ing] this kind of table, hav[ing] this kind of equipment definitely help[ed]."}
All ISLE instructors reported receiving support from community members (third panel of Fig.~\ref{MethodSupport-plot}), including ISLE co-instructors, method experts, and the broader education community. Three of the ISLE interviewees were instructors at the same institution and worked closely with one another to co-teach the introductory physics courses, forming their own community. One of these instructors (ISLE 3) expressed, \textit{``[Co-instructor 1 name], [co-instructor 2 name], and I are a team. We do everything together. And having the two of them is the most important thing."} Referring to the broader community, another ISLE instructor (ISLE 5) stated, \textit{``PER as a whole community supported me, for sure. Yeah, the teaching environment I'm in supported me, allow[ing] me... to fail, to make mistakes, and make adjustments. So, of course, the ISLE people [who gave] me an opportunity to participate in facilitating the workshop that strengthened... my understanding of how to implement ISLE"} (first row of Table~\ref{MethodSupport-table}). For the Tutorials and Peer Instruction instructors, supports for implementation were more spread across the six emergent codes (fourth and fifth panels of Fig.~\ref{MethodSupport-plot}). Community support and instructor agency were particularly salient for Peer Instruction instructors. Similar to ISLE, the former is likely partly due to two of the Peer Instruction instructors coming from the same institution and co-instructing courses together. Related to instructor agency, one Peer Instruction user (Peer Instruction 7) said, \textit{``We basically have the freedom... to design our classroom the way we like. So... I didn't have problems. I mean, clickers I asked for and got... I wasn't discouraged to do anything that I wanted to try."}

\subsubsection{Difficulties in implementation}

Overall, instructors indicated that structural challenges (e.g., constraints with classroom spaces, time, or co-instructors) and student engagement and learning challenges (e.g., lack of student buy-in) were the most frequent difficulties instructors faced in implementation (Table~\ref{MethodBarriers-table} and first panel of Fig.~\ref{MethodBarriers-plot}). 
Although unprompted, about half of the instructors mentioned difficulties related to student use of Artificial Intelligence (AI) tools and six out of the 20 instructors commented on challenges with student engagement in group work. One SCALE-UP instructor (SCALE-UP 3) reported trying to get students to engage and constantly having to remind them: \textit{```you're looking up the answer right now on ChatGPT, because you're not looking at each other or at me,'"} alluding to how the use of AI tools was affecting student engagement in discussions with their peers or the instructor. The same instructor also noted that \textit{``over the last couple of years, [it is] becoming more and more difficult to have a meaningful class discussion"} because \textit{``there are students who continue to opt out of participation."} Other less frequently mentioned difficulties in implementing named active learning methods included challenges with using or accessing available resources for the method, administrative challenges such as teaching staff shortages or lack of funding for teaching equipment, and lack of instructor training with the method.

Across the individual methods, there were slight differences in the difficulties instructors experienced. Structural challenges were particularly prominent for ISLE instructors (third panel of Fig.~\ref{MethodBarriers-plot}). For example, several of the ISLE instructors mentioned difficulties using theater-like classrooms not conducive for active learning: \textit{``The most difficult part right now is these classrooms that are... traditional lecture halls, and trying to get a space that's set up for active learning is... my current battle"} (ISLE 3). 
Student engagement and learning habits presented the most prominent challenge among Peer Instruction instructors (fifth panel of Fig.~\ref{MethodBarriers-plot}), who noted that in some cases, students tended to opt out of group work activities. One Peer Instruction user (Peer Instruction 3) mentioned, \textit{``There are students in lecture who sit by themselves and don't want to interact with others."} 

Interestingly, of the 20 total instructors, the two who indicated no difficulties in implementation were Tutorials and Peer Instruction users (fourth and fifth panel of Fig.~\ref{MethodBarriers-plot}). 

\section{Discussion}

In this study, we interviewed 20 introductory physics and astronomy instructors across different institutions in the United States who use one of SCALE-UP, ISLE, Tutorials, or Peer Instruction in their classes. We probed them on their journeys to implementing these methods, including different factors that have been previously identified in the literature as influential to the use of active learning reforms: initial exposure to the method, instructor training, and support for and difficulties in implementation. By exploring all of these factors, we provide a holistic characterization of the users of these named methods and identify patterns of adoption and implementation that vary across the four methods. In the remainder of this section, we synthesize our findings from all of these factors into three overarching takeaways and then discuss limitations to the study.

\subsection{Engagement with DBER may be influential for implementation of named methods}

Interestingly, our sample of instructors was almost equally split between DBER researchers and non-DBER researchers. The presence of DBER researchers in our study is unsurprising, as DBER researchers are more likely to use active learning methods and participate in a DBER study than non-DBER researchers. However, the comparable presence of non-DBER researchers indicates that faculty who do not conduct physics education research are also implementing these named methods. While there is probably still an over-representation of DBER faculty in our sample relative to the entire physics faculty population, this finding is encouraging for broader adoption of these methods. In their 2014 study on the diffusion of SCALE-UP, Foote \textit{et al.}~\cite{foote2014diffusion} mentioned that this method might have been at a ``tipping point" between adoption by early users (i.e., developers and/or education researchers) and more typical faculty. We observe a similar pattern for our sample of instructors and all four named active learning methods, though larger quantitative studies are needed to validate this claim.

Still, all of the non-DBER instructors in our study were engaged with DBER to some extent (e.g., reading DBER literature and attending DBER conferences), suggesting that some level of involvement in DBER is likely important for choosing to implement named active learning methods. Similarly, all instructors indicated receiving at least some form of training for EBIPs, including graduate-level coursework in education. These patterns corroborate prior work, which has identified that some involvement with DBER work and/or the DBER community is important for instructors to be made aware of these named methods~\cite{henderson2009impact}, and point to the importance of DBER for not only the development of active learning methods, but also their dissemination. Indeed, the instructors in our study interacted with DBER and learned about their methods at different stages in their careers (e.g., as graduate TAs, postdoctoral researchers, faculty members, etc.) and through different avenues (e.g., direct engagement with the method, professional development). 

Providing, encouraging, and incentivizing opportunities for physics faculty to engage with the field of DBER throughout their career, therefore, may improve the uptake of active learning methods and help to sustain implementation~\cite{brownell2012barriers, walczyk2003use}. Prior work recommends several different avenues for such opportunities. For example, physics departments could require EBIP and/or method-specific training for all students with teaching or tutoring duties within the department, including early career scholars such as undergraduate LAs and graduate TAs~\cite{brownell2012barriers}. In addition, departments may host colloquiums or seminars by DBER researchers to inform postdocs and faculty about active learning and stimulate meaningful conversations about teaching~\cite{elliott2016improved}. Increasing the presence of DBER faculty within physics departments may also promote the use of evidence-based teaching strategies~\cite{rosen2026research}. 
Finally, fostering a departmental culture that incentivizes participation in teaching-related professional development may be helpful for connecting instructors with DBER work and people outside of their home institution, for example through external workshops and conferences~\cite{yik2022association}. 


\subsection{Rising dilemmas in the face of enduring issues}

Structural challenges (e.g., inadequate time to cover content, administrative challenges such as lack of funding for resources needed, etc.) were the most commonly identified difficulties in implementation of named methods. These challenges are similar to the ``situational barriers" identified in Ref.~\cite{henderson2007barriers} and the ``leadership, organization and technology needs" uncovered in Ref.~\cite{borte2023barriers}. Research suggests that these difficulties may hinder the implementation of active learning strategies~\cite{henderson2007barriers}, but for the instructors in our study, these difficulties did not completely deter their implementation. One possible explanation for this distinction could be that our instructors reported more supports than difficulties (Figs.~\ref{MethodSupport-plot} and~\ref{MethodBarriers-plot}). Indeed, Bathgate \textit{et al.}~\cite{bathgate2019perceived} found that enhanced support might be more important for implementation of active learning strategies than attempts to eliminate all difficulties instructors face. Administrative and other forms of support to overcome structural challenges are likely crucial for sustained implementation of active learning methods; however, we encourage future research to further examine the ways in which active learning implementers combat these well-documented ``barriers."

In addition to these long-standing issues, instructors are contending with nascent issues in the active learning physics classroom related to emergent technologies. Many instructors in our study experienced recent challenges with getting students to interact with their peers and/or with classroom activities, which are at the core of active learning. Relatedly, instructors mentioned difficulties with student use of AI tools, with some noting that student use of these tools was replacing peer-to-peer and/or student-instructor interactions. 
The rise of AI, therefore, poses new hurdles that active learning instructors in particular will have to either overcome or circumvent to preserve the interactive engagement that is central to knowledge construction. 
While the use of AI tools in physics instruction is a rapidly growing area of research~\cite{kortemeyer2024cheat, sirnoorkar2026introductory, tong2025exploring, wattanakasiwich2025physics}, the role AI plays within the context of active learning environments is yet to be explored. Future research, for example using classroom observations and interviews, should directly examine student and instructor social engagement during active learning courses and their perspectives on the role of AI in engagement within active learning environments to inform new strategies for faculty to best utilize AI tools in this context.


\subsection{Emergent differences across named methods}

Although our goal was not to make direct comparisons across the four active learning methods analyzed in this study, we noticed differences across the methods with regard to method-specific training and the supports instructors received in implementing their named methods, offering several method-specific implications. 

\subsubsection{Method-specific training}

We observed differences in method-specific training and initial exposure across the four methods, which we hypothesize may be attributed to either the nature of the methods themselves and/or the resources available to instructors. Contrary to prior research~\cite{foote2014diffusion}, none of the SCALE-UP instructors reported receiving formal training to use the method, and none of them learned about the method through professional development avenues (e.g., seminars or workshops). We speculate that this may be because many of the SCALE-UP instructors in our study are newer users of the method (i.e., having taught with SCALE-UP for 5 to 10 years) and training opportunities may not be as prevalent as when the method was initially introduced. 
Beyond the instructors in our sample, a lack of formal training opportunities may be a barrier for other instructors who have learned about the method and want to implement it, as training has been shown to equip instructors for implementation~\cite{yik2022association,foote2014diffusion}. However, the SCALE-UP instructors we interviewed identified direct engagement and informal training avenues as useful for their implementation, pointing to a possible shift in the ways in which this method is disseminated. These forms of training, which are more immersive and personal (because it typically involves another user of the method), may be particularly useful for SCALE-UP users because the method requires both a new classroom layout and course structure~\cite{dancy2024physics}. Method developers and education researchers should consider new avenues to offer direct engagement experiences (e.g., observing live or video-recording class sessions) or informal training opportunities (e.g., journal clubs, community discussions) to promote SCALE-UP adoption.

On the other hand, ISLE and Tutorials instructors reported receiving multiple types of training for their method. It is noteworthy that these are also the two methods for which lack of instructor training was not mentioned as a difficulty in implementation. For ISLE in particular, instructors primarily learned about the method through direct engagement with the method, but they still participated in other forms of training beyond direct engagement. This finding suggests that ISLE instructors may need more training before implementing the method, potentially due to the method requiring a different, experimentation-oriented curriculum. Alternatively or in addition, this finding may be reflective of the many available training resources for ISLE, such as workshops, an active Facebook group, the Active Learning Guide textbook~\cite{vanheuvelen2006active} and online worksheets~\cite{ISLEweb}. Future research may explore if either or both of these possibilities explain our results. If several forms of training are necessary to support ISLE implementation, education researchers should aim to identify which aspects of each training mechanism are the most effective in order to create one holistic training environment to aid adoption. If the use of available resources is key, then ISLE developers may work to disseminate them further and developers of other methods may work to create similar types of resources to support implementation.

Tutorials and Peer Instruction instructors received training for their method primarily through professional development opportunities. For Peer Instruction, professional development is both the primary method-specific training avenue and the primary means of exposure to the method. This finding is different from Ref.~\cite{dancy2016faculty}, where the majority of the interviewed Peer Instruction users were initially exposed to the method through informal avenues and later learned more about the method through other avenues, including formal professional development. Such a difference could be because the Peer Instruction users in our study are all from graduate degree-granting institutions and thus may have had more engagement with DBER and/or DBER-related professional development opportunities than the instructors in their study, who were mostly from Bachelor's or Associate's degree-granting institutions. Alternatively, this pattern may indicate a shift in method dissemination over the past decade, with formal opportunities playing a larger role than informal avenues in promoting awareness of and providing training for Peer Instruction in recent years. Notably, the nature of both Tutorials and Peer Instruction is such that they can be readily added to a traditional, lecture-style class~\cite{dancy2024physics} (i.e., by incorporating clicker questions or worksheet activities into the lectures and/or recitations) without overhauling the entire curriculum and/or creating new classroom spaces. Therefore, attending an introductory seminar and/or workshop that provides some basic training may be sufficient for instructors to adopt one of these two methods. More immersive forms of training (e.g., direct engagement) seem to be helpful for other methods requiring a stronger pivot from traditional teaching practices, such as SCALE-UP and ISLE. 




\subsubsection{Supports for implementation}

There were also a few noteworthy differences in the supports instructors received during their implementation of named methods, which may be explained by the nature of the methods. For example, SCALE-UP users often described administrative support. This outcome is unsurprising, as SCALE-UP typically requires an unconventional classroom space and restructuring of the course to de-compartmentalize lecture, lab, and recitation. This finding is also similar to that of Ref.~\cite{foote2016enabling}, where 20 of the 21 SCALE-UP implementers they interviewed identified administrative support as a factor that helped their implementation. 
We also observed that all ISLE instructors received community support. This may be due to three of the instructors in our study working in the same department at the same institution, forming a community to co-teach the introductory physics courses. Alternatively or in addition, all of the ISLE instructors also reported receiving training through direct engagement, such as working alongside a mentor, supervisor, or other instructors. The salience of administrative support for SCALE-UP users and the importance of community for ISLE implementers highlight that although some supports are relevant across all methods, certain supports may be more beneficial, and possibly necessary, for the implementation of specific methods. Rather than broadly classifying supports in active learning, therefore, future studies should hone in on the types of supports that might be more relevant for certain active learning strategies. Future work may also identify the relationships of different supports to one another within each method, in line with the findings of Ref.~\cite{bathgate2019perceived}, and more closely map supports to when in the adoption timeline they were most useful, whether during initial exposure, early implementation, or sustained use. These lines of work will inform specific ways for developers and researchers to disseminate and support different forms of active learning to a broader set of faculty members.

\subsection{Limitations}

Our findings should be considered exploratory in light of the following limitations. First, as a qualitative study, we relied on self-reports of faculty on the different aspects of implementation we aimed to investigate, which may have been prone to recall bias. That is, what instructors described in their responses may reflect only their most memorable and/or most recent experiences, possibly leading to under-reporting. We hedged against this possibility by sending the protocol to instructors two days prior to the interview, giving them time to think about, look up, and/or recall their past experiences, increasing the accuracy of our findings. Furthermore, we asked about the teaching practices they used and the supports and difficulties they experienced for their named methods in general, rather than for a specific course or semester, reducing recall bias for specific events and/or events that happened a long time ago.

Additionally, qualitative work of this nature may be prone to social desirability bias, where the interviewee shapes their answers to appear favorable to what they believe the interviewer expects. In this study, however, we probe instructors' specific teaching experiences, which are largely value-neutral. We also gave interviewees the option to not respond to any of the interview prompts that they did not feel comfortable with (e.g., if they felt that their experiences were undesirable), further minimizing the risk of this bias impacting our work. None of the interviewees opted out of answering any of the interview prompts. 

Another possible limitation of our study is sampling bias, which could have been introduced in three different ways. First, the participants self-reported using one of the four named methods we aimed to analyze, but it is possible that their implementation significantly differed from that which was intended by the method developers. That is, the method users in our sample may not be high-fidelity implementers, in which case the themes and patterns we identified may not actually be characteristic of the named methods. We believe this possibility is unlikely, however, because our previous work~\cite{bukola2025measuring} found that most of these instructors used similar in-class activities to those of high-fidelity implementers of the same method. Second, the 20 instructors we interviewed were a subset of the 31 instructors recruited from our previous project~\cite{sundstrom2026relative}. Therefore, our sample is non-random and may primarily include instructors who are established users of their method and/or more willing to share about their teaching experiences than the broader population of physics faculty. The themes we identified may not be representative of this broader population. Third, some of the instructors (i.e., for ISLE and Peer Instruction) in our study are from the same institution. The patterns we observed for these instructors may be reflective of shared institutional contexts rather than the methods themselves. Overall, our findings and interpretations are limited to this set of instructors and may differ for other users of these four named methods or instructors who use active learning practices more broadly. Future work may investigate variation in the experiences of different faculty populations to validate or improve upon our findings. 

Finally, our comparison of the patterns across methods is limited by the small sample size of instructors using each method. Future work including a larger and more diverse sample for each method may confirm whether these patterns are consistent or simply descriptive of our specific sample of instructors. For example, our interview prompts and the emerging codes may be used to develop close-ended surveys that can be administered to a wider range of instructors who use active learning methods to further investigate our identified patterns.

\section{Conclusions}
In this work, we have adopted a holistic approach to investigating the adoption pathways of 20 physics instructors who use named active learning methods (SCALE-UP, ISLE, Tutorials, or Peer Instruction), capturing factors such as training they received, engagement with DBER, and supports and difficulties they faced in implementation. We find that for the instructors in our study, some amount of engagement with DBER, whether reading DBER literature or fully conducting DBER studies, is associated with knowledge and use of named active learning methods in physics. We also observe that in addition to long-standing issues associated with the use of active learning, such as lack of funding and administrative support available for course reforms, instructors are facing new challenges in the contemporary classroom: students are interacting less with their peers and instructors and have increased their use of new technology such as AI. Finally, we uncover several method-specific differences: SCALE-UP and ISLE may require more opportunities for professional development than Tutorials and Peer Instruction and they draw on unique forms of support (i.e., administrative for SCALE-UP and community-based for ISLE). We hope that our findings are informative for method developers, professional developers, and physics education researchers to improve the dissemination and sustained use of these particular methods. Future studies should validate or improve upon the results we find here for a larger and more diverse set of instructors, and over time as emergent technologies continue to shape the active learning classroom. 

\section*{Acknowledgments}

We thank the instructors who participated in our study. We also thank David Esparza and Ian Olivant for their feedback on this manuscript. This material is based upon work supported by the National Science Foundation under Grant Nos. 2111128 and 2111275 and by an American Physical Society Forum on Education Project Mini-Grant.

\section*{Data Availability}

The data that support the findings of this article are not publicly available because they contain sensitive personal information. The data are available from the authors upon reasonable request.

\bibliography{bibfile}

@article{wieman2013use,
  title={Use of research-based instructional strategies: {How} to avoid faculty quitting},
  author={Wieman, Carl and Deslauriers, Louis and Gilley, Brett},
  journal={{Physical Review Special Topics-Physics Education Research}},
  volume={9},
  number={2},
  pages={023102},
  year={2013},
  publisher={APS}
}

@inproceedings{yik2022association,
  title={Association of malleable factors with adoption of research-based instructional strategies in introductory chemistry, mathematics, and physics},
  author={Yik, Brandon J. and Raker, Jeffrey R. and Apkarian, Naneh and Stains, Marilyne and Henderson, Charles and Dancy, Melissa H. and Johnson, Estrella},
  booktitle={Frontiers in Education},
  volume={7},
  pages={1016415},
  year={2022},
  organization={Frontiers Media SA}
}

@article{lund2015importance,
  title={The importance of context: an exploration of factors influencing the adoption of student-centered teaching among chemistry, biology, and physics faculty},
  author={Lund, Travis J. and Stains, Marilyne},
  journal={{International Journal of STEM Education}},
  volume={2},
  number={1},
  pages={13},
  year={2015},
  publisher={Springer}
}

@article{bathgate2019perceived,
  title={Perceived supports and evidence-based teaching in college {STEM}},
  author={Bathgate, Meghan E. and Arag{\'o}n, Oriana R. and Cavanagh, Andrew J. and Waterhouse, Jonathan K. and Frederick, Jennifer and Graham, Mark J.},
  journal={{International Journal of STEM Education}},
  volume={6},
  number={1},
  pages={11},
  year={2019},
  publisher={Springer}
}

@article{borte2023barriers,
  title={Barriers to student active learning in higher education},
  author={B{\o}rte, Kristin and Nesje, Katrine and Lillejord, S{\o}lvi},
  journal={{Teaching in Higher Education}},
  volume={28},
  number={3},
  pages={597--615},
  year={2023},
  publisher={Taylor \& Francis}
}

@article{freeman2014active,
  title={Active learning increases student performance in science, engineering, and mathematics},
  author={Freeman, Scott and Eddy, Sarah L. and McDonough, Miles and Smith, Michelle K. and Okoroafor, Nnadozie and Jordt, Hannah and Wenderoth, Mary Pat},
  journal={{Proceedings of the National Academy of Sciences}},
  volume={111},
  number={23},
  pages={8410--8415},
  year={2014},
  publisher={National Academy of Sciences},
url={https://doi.org/10.1073/pnas.1319030111}
}

@article{theobald2020active,
  title={Active learning narrows achievement gaps for underrepresented students in undergraduate science, technology, engineering, and math},
  author={Theobald, Elli J. and Hill, Mariah J. and Tran, Elisa and Agrawal, Sweta and Arroyo, E. Nicole and Behling, Shawn and Chambwe, Nyasha and Cintr{\'o}n, Dianne Laboy and Cooper, Jacob D. and Dunster, Gideon and others},
  journal={{Proceedings of the National Academy of Sciences}},
  volume={117},
  number={12},
  pages={6476--6483},
  year={2020},
  publisher={National Academy of Sciences}
}

@inproceedings{bullard2008effects,
  title={Effects of active learning on student performance and retention},
  author={Bullard, Lisa and Felder, Richard and Raubenheimer, Dianne},
  booktitle={2008 Annual Conference \& Exposition},
  pages={13--473},
  year={2008}
}

@article{salomone2017required,
  title={Required peer-cooperative learning improves retention of {STEM} majors},
  author={Salomone, Matthew and Kling, Thomas},
  journal={{International Journal of STEM Education}},
  volume={4},
  number={1},
  pages={19},
  year={2017},
  publisher={Springer}
}

@article{stump2014engineering,
  title={Engineering students' intelligence beliefs and learning},
  author={Stump, Glenda S. and Husman, Jenefer and Corby, Marcia},
  journal={{Journal of Engineering Education}},
  volume={103},
  number={3},
  pages={369--387},
  year={2014},
  publisher={Wiley Online Library}
}

@article{lumpkin2015student,
  title={Student perceptions of active learning.},
  author={Lumpkin, Angela L. and Achen, Rebecca M. and Dodd, Regan K.},
  journal={{College Student Journal}},
  volume={49},
  number={1},
  pages={121},
  year={2015}
}

@article{olson2012engage,
  title={Engage to excel: producing one million additional college graduates with degrees in science, technology, engineering, and mathematics. {Report} to the president.},
  author={Olson, Steve and Riordan, Donna Gerardi},
  journal={{Executive Office of the President}},
  year={2012},
  publisher={ERIC}
}

@book{mazur1997peer,
  title={{Peer Instruction}: {A} User's Manual},
  author={Mazur, Eric},
  year={1997},
  publisher={Prentice Hall}
}

@article{hake1998interactive,
  title={Interactive-engagement versus traditional methods: {A} six-thousand-student survey of mechanics test data for introductory physics courses},
  author={Hake, Richard R.},
  journal={{American Journal of Physics}},
  volume={66},
  number={1},
  pages={64--74},
  year={1998},
  publisher={American Association of Physics Teachers}
}

@article{stains2018anatomy,
  title={Anatomy of {STEM} teaching in {N}orth {A}merican universities},
  author={Stains, Marilyne and Harshman, Jordan and Barker, Megan K. and Chasteen, Stephanie V. and Cole, Renee and DeChenne-Peters, Sue Ellen and Eagan Jr, M. Kevin and Esson, Joan M. and Knight, Jennifer K. and Laski, Frank A. and others},
  journal={{Science}},
  volume={359},
  number={6383},
  pages={1468--1470},
  year={2018},
  publisher={American Association for the Advancement of Science}
}

@article{henderson2007barriers,
  title={Barriers to the use of research-based instructional strategies: The influence of both individual and situational characteristics},
  author={Henderson, Charles and Dancy, Melissa H.},
  journal={{Physical Review Special Topics-Physics Education Research}},
  volume={3},
  number={2},
  pages={020102},
  year={2007},
  publisher={APS}
}

@article{albuquerque2023awareness,
  title={Awareness and adoption of evidence-based instructional practices by {STEM} faculty in the {UAE} and {USA}},
  author={Albuquerque, Melinda Joy Biggs and Awadalla, Dina Mustafa Mohammad and de Albuquerque, Francisco Daniel Benicio and Aly Hassan, Ashraf},
  journal={{Education Sciences}},
  volume={13},
  number={2},
  pages={204},
  year={2023},
  publisher={MDPI}
}

@article{yik2022evaluating,
  title={Evaluating the impact of malleable factors on percent time lecturing in gateway chemistry, mathematics, and physics courses},
  author={Yik, Brandon J. and Raker, Jeffrey R. and Apkarian, Naneh and Stains, Marilyne and Henderson, Charles and Dancy, Melissa H. and Johnson, Estrella},
  journal={{International Journal of STEM Education}},
  volume={9},
  number={1},
  pages={15},
  year={2022},
  publisher={Springer}
}

@article{henderson2012use,
  title={Use of research-based instructional strategies in introductory physics: Where do faculty leave the innovation-decision process?},
  author={Henderson, Charles and Dancy, Melissa and Niewiadomska-Bugaj, Magdalena},
  journal={{Physical Review Special Topics-Physics Education Research}},
  volume={8},
  number={2},
  pages={020104},
  year={2012},
  publisher={APS}
}

@article{gardner2021exploring,
  title={Exploring barriers to the use of evidence-based instructional practices},
  author={Gardner, Grant E. and Brown, Evelyn and Grimes, Zachary and Bishara, Gina},
  journal={{Journal of College Science Teaching}},
  volume={51},
  number={2},
  pages={56--66},
  year={2021},
  publisher={Taylor \& Francis}
}

@article{beichner2007student,
  title={The student-centered activities for large enrollment undergraduate programs {(SCALE-UP)} project},
  author={Beichner, Robert J. and Saul, Jeffery M. and Abbott, David S. and Morse, Jeanne J. and Deardorff, Duane and Allain, Rhett J. and Bonham, Scott W. and Dancy, Melissa H. and Risley, John S.},
  journal={{Research-Based Reform of University Physics}},
  volume={1},
  number={1},
  pages={2--39},
  year={2007},
  publisher={American Association of Physics Teachers College Park, MD}, 
url={https://doi.org/10.1119/RevPERv1.1.4}
}

@inproceedings{beichner2000introduction,
  title={Introduction to {SCALE-UP}: {Student-Centered Activities for Large Enrollment University Physics}},
  author={Beichner, Robert J. and Saul, Jeffery M. and Allain, Rhett J. and Deardorff, Duane L. and Abbott, David S.},
  year={2000},
  publisher={ERIC},
  note={Paper presented at the Annual Meeting of the American Association for Engineering Education}
}

@inproceedings{beichner1999student,
  title={Student-centered activities for large-enrollment university physics {(SCALE-UP)}},
  author={Beichner, Robert J. and Saul, J.},
  booktitle={Proceedings of the Sigma Xi Forum on the Reform of Undergraduate Education},
  pages={43--52},
  year={1999}
}

@techreport{beichner2008scale,
  title={The {SCALE-UP} {Project}: {A} student-centered active learning environment for undergraduate programs},
  author={Beichner, Robert},
  institution={National Academy of Sciences},
  year={2008}
}

@article{etkina2007investigative,
  title={Investigative science learning environment--{A} science process approach to learning physics},
  author={Etkina, Eugenia and Van Heuvelen, Alan and others},
  journal={{Research-Based Reform of University Physics}},
  volume={1},
  number={1},
  pages={1--48},
  year={2007},
  publisher={American Association of Physics Teachers College Park, MD},
url={https://doi.org/10.1119/RevPERv1.1.1}
}

@inproceedings{etkina2021investigative,
  title={The investigative science learning environment {(ISLE)} approach to learning physics},
  author={Etkina, E. and Brookes, D. T. and Planinsic, G.},
  booktitle={Journal of Physics: Conference Series},
  volume={1882},
  number={1},
  pages={012001},
  year={2021},
  organization={IOP Publishing}
}

@article{etkina2006using,
  title={Using introductory labs to engage students in experimental design},
  author={Etkina, Eugenia and Murthy, Sahana and Zou, Xueli},
  journal={{American Journal of Physics}},
  volume={74},
  number={11},
  pages={979--986},
  year={2006},
  publisher={American Association of Physics Teachers}
}

@article{finkelstein2005replicating,
  title={Replicating and understanding successful innovations: Implementing tutorials in introductory physics},
  author={Finkelstein, Noah D. and Pollock, Steven J.},
  journal={{Physical Review Special Topics-Physics Education Research}},
  volume={1},
  number={1},
  pages={010101},
  year={2005},
  publisher={APS}
}

@book{adams2003lecture,
  title={{Lecture-Tutorials for Introductory Astronomy}},
  author={Adams, J. P. and Prather, E. E. and Slater, T. F.},
  year={2005},
publisher={Prentice Hall, Upper Saddle River, NJ}
}

@article{crouch2001peer,
  title={Peer instruction: {Ten} years of experience and results},
  author={Crouch, Catherine H. and Mazur, Eric},
  journal={{American Journal of Physics}},
  volume={69},
  number={9},
  pages={970--977},
  year={2001},
  publisher={American Association of Physics Teachers}
}

@inproceedings{mazur1997peerconf,
  title={Peer instruction: {Getting} students to think in class},
  author={Mazur, Eric},
  booktitle={AIP conference proceedings},
  pages={981--988},
  year={1997},
  organization={IOP Institute Of Physics Publishing Ltd.}
}

@article{missett2015searching,
  title={Searching for evidence-based practice: {A} survey of empirical studies on curricular interventions measuring and reporting fidelity of implementation published during 2004-2013},
  author={Missett, Tracy C. and Foster, Lisa H.},
  journal={{Journal of Advanced Academics}},
  volume={26},
  number={2},
  pages={96--111},
  year={2015},
  publisher={SAGE Publications Sage CA: Los Angeles, CA}
}

@article{rahman2020evaluating,
  title={Evaluating the evidence base for evidence-based instructional practices in chemistry through meta-analysis},
  author={Rahman, Tawabur and Lewis, Scott E.},
  journal={{Journal of Research in Science Teaching}},
  volume={57},
  number={5},
  pages={765--793},
  year={2020},
  publisher={Wiley Online Library}
}

@article{dancy2024physics,
  title={Physics instructors’ knowledge and use of active learning has increased over the last decade but most still lecture too much},
  author={Dancy, Melissa and Henderson, Charles and Apkarian, Naneh and Johnson, Estrella and Stains, Marilyne and Raker, Jeffrey R. and Lau, Alexandra},
  journal={{Physical Review Physics Education Research}},
  volume={20},
  number={1},
  pages={010119},
  year={2024},
  publisher={APS}
}

@article{walczyk2003use,
  title={Use of learner-centered instruction in college science and mathematics classrooms},
  author={Walczyk, Jeffrey J. and Ramsey, Linda L.},
  journal={{Journal of Research in Science Teaching: The Official Journal of the National Association for Research in Science Teaching}},
  volume={40},
  number={6},
  pages={566--584},
  year={2003},
  publisher={Wiley Online Library}
}

@article{brawner2002survey,
  title={A survey of faculty teaching practices and involvement in faculty development activities},
  author={Brawner, Catherine E. and Felder, Richard M. and Allen, Rodney and Brent, Rebecca},
  journal={{Journal of Engineering Education}},
  volume={91},
  number={4},
  pages={393--396},
  year={2002},
  publisher={Wiley Online Library}
}

@article{henderson2008promoting,
  title={Promoting instructional change in new faculty: {An} evaluation of the physics and astronomy new faculty workshop},
  author={Henderson, Charles},
  journal={{American Journal of Physics}},
  volume={76},
  number={2},
  pages={179--187},
  year={2008},
  publisher={AIP Publishing}
}

@article{henderson2011facilitating,
  title={Facilitating change in undergraduate {STEM} instructional practices: {An} analytic review of the literature},
  author={Henderson, Charles and Beach, Andrea and Finkelstein, Noah},
  journal={{Journal of Research in Science Teaching}},
  volume={48},
  number={8},
  pages={952--984},
  year={2011},
  publisher={Wiley Online Library}
}

@article{andrews2011active,
  title={Active learning not associated with student learning in a random sample of college biology courses},
  author={Andrews, Tessa M. and Leonard, Michael J. and Colgrove, Clinton A. and Kalinowski, Steven T.},
  journal={{CBE-Life Sciences Education}},
  volume={10},
  number={4},
  pages={394--405},
  year={2011},
  publisher={American Society for Cell Biology},
url={https://doi.org/10.1187/cbe.11-07-0061}
}

@incollection{sokoloff2021active,
  title={Active dissemination - {Over} three decades of faculty development in active learning},
  author={Sokoloff, David R.},
  booktitle={Teaching-Learning Contemporary Physics: From Research to Practice},
  pages={201--211},
  year={2021},
  publisher={Springer}
}

@article{rossi2021active,
  title={Active learning tools improve the learning outcomes, scientific attitude, and critical thinking in higher education: {Experiences} in an online course during the {COVID-19} pandemic},
  author={Rossi, Izadora Volpato and de Lima, Jordana Dinor{\'a} and Sabatke, Bruna and Nunes, Maria Alice Ferreira and Ramirez, Graciela Evans and Ramirez, Marcel Ivan},
  journal={{Biochemistry and Molecular Biology Education}},
  volume={49},
  number={6},
  pages={888--903},
  year={2021},
  publisher={Wiley Online Library}
}

@article{dancy2016faculty,
  title={How faculty learn about and implement research-based instructional strategies: {The} case of peer instruction},
  author={Dancy, Melissa and Henderson, Charles and Turpen, Chandra},
  journal={{Physical Review Physics Education Research}},
  volume={12},
  number={1},
  pages={010110},
  year={2016},
  publisher={APS}
}

@article{commeford2021characterizing,
  title={Characterizing active learning environments in physics using network analysis and classroom observations},
  author={Commeford, Kelley and Brewe, Eric and Traxler, Adrienne},
  journal={{Physical Review Physics Education Research}},
  volume={17},
  number={2},
  pages={020136},
  year={2021},
  publisher={APS}
}

@article{commeford2022characterizing,
  title={Characterizing active learning environments in physics using latent profile analysis},
  author={Commeford, Kelley and Brewe, Eric and Traxler, Adrienne},
  journal={{Physical Review Physics Education Research}},
  volume={18},
  number={1},
  pages={010113},
  year={2022},
  publisher={APS}
}

@article{turpen2009not,
  title={Not all interactive engagement is the same: {Variations} in physics professors’ implementation of {Peer Instruction}},
  author={Turpen, Chandra and Finkelstein, Noah D.},
  journal={{Physical Review Special Topics-Physics Education Research}},
  volume={5},
  number={2},
  pages={020101},
  year={2009},
  publisher={APS}
}

@book{mcdermott2002tutorials,
  title={Tutorials in {I}ntroductory {P}hysics},
  author={McDermott, Lillian C.},
  year={2002},
  publisher={Prentice Hall}
}

@article{Affriyenni2025navigating,
  title = {Navigating the adoption of research-based instructional strategies within the complex nature of higher education},
  author = {Affriyenni, Yessi and Georgiou, Helen and Finkelstein, Noah},
  journal = {{Physical Review Physics Education Research}},
  volume = {21},
  issue = {2},
  pages = {020124},
  numpages = {21},
  year = {2025},
  month = {Sep},
  publisher = {American Physical Society},
  doi = {10.1103/wvjk-8y8k},
  url = {https://link.aps.org/doi/10.1103/wvjk-8y8k}
}

@article{heller1992teaching,
  title={Teaching problem solving through cooperative grouping},
  author={Heller, Patricia and Keith, Ronald and Anderson, Scott},
  journal={{American Journal of Physics}},
  volume={60},
  number={7},
  pages={627--636},
  year={1992}
}

@article{heller1992teaching2,
  title={Teaching problem solving through cooperative grouping. {Part 2}: {Designing} problems and structuring groups},
  author={Heller, Patricia and Hollabaugh, Mark},
  journal={{American Journal of Physics}},
  volume={60},
  number={7},
  pages={637--644},
  year={1992},
  publisher={AIP Publishing}
}

@article{brewe2008modeling,
  title={Modeling theory applied: {Modeling Instruction} in introductory physics},
  author={Brewe, Eric},
  journal={{American Journal of Physics}},
  volume={76},
  number={12},
  pages={1155--1160},
  year={2008},
  publisher={AIP Publishing}
}

@article{henderson2009impact,
  title={Impact of physics education research on the teaching of introductory quantitative physics in the {United States}},
  author={Henderson, Charles and Dancy, Melissa H.},
  journal={{Physical Review Special Topics-Physics Education Research}},
  volume={5},
  number={2},
  pages={020107},
  year={2009},
  publisher={APS}
}

@article{sundstrom2026beyond,
  title={Beyond named methods: {A} typology of active learning based on classroom observation networks},
  author={Sundstrom, Meagan and Gambrell, Justin and Green, Colin and Traxler, Adrienne L. and Brewe, Eric},
  journal={{Physical Review Physics Education Research}},
  volume={22},
  number={1},
  pages={010143},
  year={2026},
  publisher={APS}
}

@article{foote2014diffusion,
  title={Diffusion of research-based instructional strategies: {The} case of {SCALE-UP}},
  author={Foote, Kathleen T. and Neumeyer, Xaver and Henderson, Charles and Dancy, Melissa H. and Beichner, Robert J.},
  journal={{International Journal of STEM Education}},
  volume={1},
  number={1},
  pages={10},
  year={2014},
  publisher={Springer}
}

@inproceedings{bukola2025perc,
author = "Ibukunoluwa Bukola and Meagan Sundstrom and Justin Gambrell and Olive Ross and Adrienne Traxler and Eric Brewe", 
title = {Multi-institutional assessment of {Peer Instruction} implementation and impacts using the {Framework for Interactive Learning in Lectures}}, booktitle = {Physics Education Research Conference 2025}, 
pages = {105-111}, 
address = {Washington, DC}, 
series = {PER Conference}, 
month = {August 6-7}, 
year = {2025} 
}

@inproceedings{sundstrom2025evaluating,
author = {Meagan Sundstrom and Justin Gambrell and Adrienne Traxler and Eric Brewe}, 
title = {Evaluating recognition and recall formats of social network surveys in physics education research}, 
booktitle = {Physics Education Research Conference 2025}, 
pages = {417-422}, 
address = {Washington, DC}, 
Series = {PER Conference}, 
month = {August 6-7}, 
year = {2025}
}

@inproceedings{hall2002adoption,
  title={Adoption of active learning in a lecture-based engineering class},
  author={Hall, Steven R. and Waitz, Ian and Brodeur, Doris R. and Soderholm, Diane H. and Nasr, Reem},
  booktitle={32nd Annual frontiers in education},
  volume={1},
  pages={T2A--T2A},
  year={2002},
  organization={IEEE}
}

@inproceedings{turpen2007understanding,
  title={Understanding how physics faculty use peer instruction},
  author={Turpen, Chandra and Finkelstein, Noah D.},
  booktitle={AIP conference proceedings},
  volume={951},
  number={1},
  pages={204--207},
  year={2007},
  organization={American Institute of Physics}
}

@article{chasteen2015educational,
  title={Educational transformation in upper-division physics: {The Science Education Initiative} model, outcomes, and lessons learned},
  author={Chasteen, Stephanie V. and Wilcox, Bethany and Caballero, Marcos D. and Perkins, Katherine K. and Pollock, Steven J. and Wieman, Carl E.},
  journal={{Physical Review Special Topics-Physics Education Research}},
  volume={11},
  number={2},
  pages={020110},
  year={2015},
  publisher={APS}
}

@article{givens2020faculty,
  title={{Faculty Perceptions of the Implementation of Flipped Teaching in Undergraduate STEM Courses Across Two Institutions}},
  author={Givens, Adriana and Bracey, Georgia and Fickas, Julie and Bartels, Lynn and Locke, Sharon and Gopalan, Chaya},
  journal={{The FASEB Journal}},
  volume={34},
  number={S1},
  pages={1--1},
  year={2020},
  publisher={Wiley Online Library}
}

@article{west2022stem,
  title={{STEM} faculty institute: an intensive interdisciplinary effort to improve {STEM} faculty adoption of evidence-based instructional practices},
  author={West, Richard E. and Jensen, Jamie L. and Johnson, Michael and Nielson, Jennifer and Sansom, Rebecca and Wright, Geoffrey},
  journal={{Journal of College Science Teaching}},
  volume={51},
  number={3},
  pages={78--87},
  year={2022},
  publisher={Taylor \& Francis}
}

@article{funk2022role,
  title={The role of leadership in educational innovation: a comparison of two mathematics departments’ initiation, implementation, and sustainment of active learning},
  author={Funk, Rachel and Uhing, Karina and Williams, Molly and Smith, Wendy M.},
  journal={{SN Social Sciences}},
  volume={2},
  number={12},
  pages={258},
  year={2022},
  publisher={Springer}
}

@article{foote2016enabling,
  title={Enabling and challenging factors in institutional reform: {The} case of {SCALE-UP}},
  author={Foote, Kathleen and Knaub, Alexis and Henderson, Charles and Dancy, Melissa and Beichner, Robert J.},
  journal={{Physical Review Physics Education Research}},
  volume={12},
  number={1},
  pages={010103},
  year={2016},
  publisher={APS}
}

@article{brownell2012barriers,
  title={Barriers to faculty pedagogical change: {Lack} of training, time, incentives, and... tensions with professional identity?},
  author={Brownell, Sara E. and Tanner, Kimberly D.},
  journal={{CBE-Life Sciences Education}},
  volume={11},
  number={4},
  pages={339--346},
  year={2012},
  publisher={American Society for Cell Biology}
}

@article{elliott2016improved,
  title={Improved student learning through a faculty learning community: {How} faculty collaboration transformed a large-enrollment course from lecture to student centered},
  author={Elliott, Emily R. and Reason, Robert D. and Coffman, Clark R. and Gangloff, Eric J. and Raker, Jeffrey R. and Powell-Coffman, Jo Anne and Ogilvie, Craig A.},
  journal={{CBE-Life Sciences Education}},
  volume={15},
  number={2},
  pages={ar22},
  year={2016},
  publisher={American Society for Cell Biology}
}

@article{rosen2026research,
  title={Research on physics education where physics education happens},
  author={Rosen, Drew J. and Heron, Paula R. L. and Holmes, N. G.},
  journal={{Nature Physics}},
  volume={22},
  number={6},
  pages={814--815},
  year={2026},
  publisher={Nature Publishing Group}
}

@article{bathgate2019supports,
  title={Supports: {A} key factor in faculty implementation of evidence-based teaching},
  author={Bathgate, Meghan E. and Arag{\'o}n, Oriana R. and Cavanagh, Andrew J. and Frederick, Jennifer and Graham, Mark J.},
  journal={{CBE-Life Sciences Education}},
  volume={18},
  number={2},
  pages={ar22},
  year={2019},
  publisher={American Society for Cell Biology}
}

@article{kortemeyer2024cheat,
  title={Cheat sites and artificial intelligence usage in online introductory physics courses: {What} is the extent and what effect does it have on assessments?},
  author={Kortemeyer, Gerd and Bauer, Wolfgang},
  journal={{Physical Review Physics Education Research}},
  volume={20},
  number={1},
  pages={010145},
  year={2024},
  publisher={APS}
}

@article{sirnoorkar2026introductory,
  title={Introductory physics students’ valued features in {AI} feedback generated from self-crafted and engineered prompts},
  author={Sirnoorkar, Amogh and Rebello, N. Sanjay},
  journal={{Physical Review Physics Education Research}},
  volume={22},
  number={1},
  pages={010146},
  year={2026},
  publisher={APS}
}

@article{tong2025exploring,
  title={Exploring the role of human-{AI} collaboration in solving scientific problems},
  author={Tong, Dazhen and Jin, Bangjian and Tao, Yang and Ren, Hongmei and Atiquil Islam, A. Y. M. and Bao, Lei},
  journal={{Physical Review Physics Education Research}},
  volume={21},
  number={1},
  pages={010149},
  year={2025},
  publisher={APS}
}

@article{wattanakasiwich2025physics,
  title={Physics instructors’ acceptance and implementation of generative {AI}},
  author={Wattanakasiwich, Pornrat and Kaewkhong, Kreetha and Katwibun, Duanghathai},
  journal={{Physical Review Physics Education Research}},
  volume={21},
  number={1},
  pages={010155},
  year={2025},
  publisher={APS}
}

@book{vanheuvelen2006active,
  author    = {Van Heuvelen, Alan and Etkina, Eugenia},
  title     = {{Active Learning Guide}},
  publisher = {Pearson Education},
  year      = {2005},
  isbn      = {9780805390780}
}

@misc{ISLEweb,
    howpublished = {\url{https://sites.google.com/site/scientificabilities/}}
}

@article{sundstrom2026relative,
  title={Relative benefits of different active learning methods to conceptual physics learning},
  author={Sundstrom, Meagan and Gambrell, Justin and Green, Colin and Traxler, Adrienne L. and Brewe, Eric},
  journal={{Nature Physics}},
  pages={1--6},
  year={2026},
  publisher={Nature Publishing Group UK London}
}

@article{bukola2025measuring,
  title={Measuring fidelity of implementation of named active learning methods in physics},
  author={Bukola, Ibukunoluwa and Sundstrom, Meagan and Gambrell, Justin and Green, Colin and Traxler, Adrienne L. and Brewe, Eric},
  journal={{Physical Review Physics Education Research}},
  volume={22},
  number={2},
  pages={020142},
  year={2026},
  publisher={APS}
}

@article{mcpadden2023planning,
  title={Planning for participants’ varying needs and abilities in qualitative research},
  author={McPadden, Daryl and Sawtelle, Vashti and Scanlon, Erin M. and Chini, Jacquelyn J. and Chahal, Harsna and Levy, Regan and Reynolds, Alex},
  journal={{Physical Review Physics Education Research}},
  volume={19},
  number={2},
  pages={020143},
  year={2023},
  publisher={APS}
}

@book{tesch2013qualitative,
  title={Qualitative research: {Analysis} types and software},
  author={Tesch, Renata},
  year={2013},
  publisher={Routledge}
}

@book{drever1995using,
  title={{Using Semi-Structured Interviews in Small-Scale Research. A Teacher's Guide.}},
  author={Drever, Eric},
  year={1995},
  publisher={ERIC}
}

@article{ruslin2022semi,
  title={Semi-structured interview: {A} methodological reflection on the development of a qualitative research instrument in educational studies},
  author={Ruslin, Ruslin and Mashuri, Saepudin and Rasak, Muhammad Sarib Abdul and Alhabsyi, Firdiansyah and Syam, Hijrah},
  journal={{IOSR Journal of Research \& Method in Education (IOSR-JRME)}},
  volume={12},
  number={1},
  pages={22--29},
  year={2022}
}

@article{campbell2013coding,
  title={Coding in-depth semistructured interviews: {Problems} of unitization and intercoder reliability and agreement},
  author={Campbell, John L. and Quincy, Charles and Osserman, Jordan and Pedersen, Ove K.},
  journal={{Sociological Methods \& Research}},
  volume={42},
  number={3},
  pages={294--320},
  year={2013},
  publisher={Sage Publications Sage CA: Los Angeles, CA}
}

@article{wilson1994cuple,
  title={The CUPLE physics studio},
  author={Wilson, Jack M},
  journal={The Physics Teacher},
  volume={32},
  number={9},
  pages={518--523},
  year={1994},
  publisher={American Association of Physics Teachers}
}

\clearpage

\section*{APPENDIX}

\subsection{Interview Protocol}
We asked all interviewees the questions listed below. The brackets indicate places where we included the method name or method-specific activities relevant to the instructor being interviewed. 
\vspace{\baselineskip}
\begin{enumerate}
    \item How long have you been teaching physics for, and in what roles (e.g., assistant professor)?
    \item Did you receive any formal training in course design, evidence-based instructional practices, or other forms of teaching, and if so, where (e.g., graduate course, workshops, Faculty Teaching Institute)?
    \item What is your experience or familiarity with discipline-based education research, if any?
    \item How and when did you first learn about [active learning method]?
    \item Did you receive any formal training on using [active learning method], and if so, where?
    \item What external resources (e.g., PhysPort) do you use to design classroom activities, if any? 
    \item How long have you been using [active learning method] in your physics classes?
    \item Why did you decide to use [active learning method] as compared to other teaching methods?
    \item What do you think are the central features or components of [active learning method]?
    \begin{enumerate}
        \item What has informed your knowledge of the central components (e.g., training, literature, conversations with colleagues)?
    \end{enumerate}
    \item We conducted a literature review to identify the central features of [active learning method]: [list of critical components of the named method, see next subsection]. 
    For each of these components:
    
    \begin{enumerate}
        \item Do you use this in your classes?
        \begin{itemize}
        \item Why or why not?
        \item What does your implementation of this component look like in practice?
        
        \end{itemize}
        \item How, if at all, does your implementation of this component differ from the original design of the method?
        \begin{itemize}
        \item What challenges or needs prompted these modifications? 
        \end{itemize}
        
    \end{enumerate}
    \item What additional components that are not part of [active learning method] (i.e., that are not included in the list of components we just discussed) do you use in your classes, if there are any?
    \begin{enumerate}
        \item Why did you decide to use these components?
    \end{enumerate}
   
    \item What has supported you in implementing [active learning method] in your classes?
    \item What has made it difficult for you to implement [active learning method] in your classes?
    \item Recent research has shown that individual students asking questions in front of the whole class benefits the learning of all students in the class, why do you think that might be?
    \item Under what circumstances do you think a student question is not a productive use of class time?
    \item In what ways, if at all, do you encourage students to ask questions in front of the whole class?
    \item Thinking back on classes you have taught where students asked a lot of questions and those where students did not ask many questions, what do you think might have caused those differences?
    \item Is there anything else you would like to share about your teaching values or use of [active learning method]?
\end{enumerate}

\subsubsection{Critical components of named methods}

For question 10 in the interview protocol, we asked about the following critical components for each method. We identified the critical components through a review of the literature for each method, as detailed in Ref.~\cite{bukola2025measuring}.\\

SCALE-UP:
\begin{itemize}
    \item Restaurant-style classrooms (e.g., tables with nine seats for three groups of students)
    \item Lab, lecture, and recitation integrated together in one course section
    \item New material introduced through preclass readings and quizzes
    \item Students work on tangibles (hands-on activities), ponderables (open-ended conceptual questions), and experiments in groups
    \item Students present their answers to the whole class
    \item Class-wide discussions after completing activities
\end{itemize}
\vspace{\baselineskip}

    ISLE:
\begin{itemize}
    \item TA training on facilitating group work and working through activities that students will be expected to cover; conducted for at least an hour weekly
    \item Students complete readings after class
    \item Students devise explanations for observation experiments in groups
    \item Students create testing experiments and predict outcomes of the experiments in groups
    \item Students solve problems and respond with a classroom response system (e.g., clickers)
\end{itemize}

\vspace{\baselineskip}
Tutorials:
\begin{itemize}
    \item TA training on content knowledge and facilitating group work (not providing answers but encouraging student reasoning), conducted weekly
    \item Students complete pretests and posttests (or homework) on in-class activities
    \item Students complete worksheets in groups
\end{itemize}

\vspace{\baselineskip}
Peer Instruction:
\begin{itemize}
    \item Students complete readings before class
    \item Students complete mini-quizzes or assignments to confirm preclass reading has been done
    \item Brief lectures or lecture demonstrations by instructor
    \item Students work individually on conceptual questions (e.g., clicker questions) and then in groups
    \item Polls of student answers to conceptual questions are displayed in front of the whole class
    \item Students receive feedback on their responses to conceptual questions
\end{itemize}

\end{document}